\documentclass[journal]{IEEEtran}
\usepackage{amsmath,amsfonts}
\usepackage{algorithmic}
\usepackage{algorithm}
\usepackage{array}
\usepackage[caption=false,font=footnotesize]{subfig} 
\usepackage{textcomp}
\usepackage{stfloats}
\usepackage{url}
\usepackage{verbatim}
\usepackage[dvipdfmx]{graphicx}
\usepackage{cite}

\usepackage{bm}
\usepackage{enumerate}
\usepackage{booktabs}
\usepackage{framed}
\usepackage{color}

\newtheorem{thm}{Theorem}
\newtheorem{lem}{Lemma}

\newtheorem{prob}{Problem}
\newtheorem{remark}{Remark}

\def\zero{\bm{0}}
\def\one{\bm{1}}

\def\a{\bm{a}}
\def\b{\bm{b}}
\def\c{\bm{c}}

\def\h{\bm{h}}

\def\p{\bm{p}}

\def\s{\bm{s}}
\def\t{\bm{t}}

\def\v{\bm{v}}
\def\w{\bm{w}}
\def\x{\bm{x}}
\def\y{\bm{y}}

\def\gammab{\bm{\gamma}}
\def\taub{\bm{\tau}}

\def\FC{\mathcal{F}}
\def\GC{\mathcal{G}}
\def\IC{\mathcal{I}}
\def\JC{\mathcal{J}}

\def\LC{\mathcal{L}}
\def\MC{\mathcal{M}}
\def\NC{\mathcal{N}}
\def\SC{\mathcal{S}}

\def\ALN{[A(\LC), A(\NC \setminus \LC)]}
\def\ALM{[A(\LC), A(\MC \setminus \LC)]}

\def\Real{\mathbb{R}}
\def\equivSym{\Leftrightarrow}
\def\trans{\top}
\def\trace{\mathrm{tr}}

\def\diag{\mathrm{diag}}

\def\opt{\mathrm{opt}}

\def\score{\mathrm{score}}
\def\diam{\mathrm{diam}}
\def\MRSA{\mathrm{MRSA}}

\def\NH{\mathrm{NH}}
\def\ave{\mathrm{ave}}
\def\real{\mathrm{real}}
\def\reference{\mathrm{ref}}

\def\HT{\mathsf{H}}
\def\Primal{\mathsf{P}}
\def\Dual{\mathsf{D}}
\def\AugProb{\mathsf{R}}
\def\OptProb{\mathsf{S}}

\title{Endmember Extraction from Hyperspectral Images Using Self-Dictionary Approach with Linear Programming}

\author{Tomohiko~Mizutani
\thanks{Department of Mathematical and Systems Engineering,
Shizuoka University,
3-5-1 Johoku, Chuo, Hamamatsu, 432-8561, Japan.
{\tt mizutani.t@shizuoka.ac.jp}}}

\date{\today}

\begin{document}

\maketitle

\begin{abstract}
    Hyperspectral imaging technology has a wide range of applications,
    including forest management, mineral resource exploration, and Earth surface monitoring.
    A key step in utilizing this technology is endmember extraction,
    which aims to identify the spectral signatures of materials in  observed scenes.
    Theoretical studies suggest that self-dictionary methods using linear programming (LP), known as Hottopixx methods,
    are effective in extracting endmembers.
    However, their practical application is hindered by high computational costs,
    as they require solving LP problems whose size grows quadratically with the number of pixels in the image.
    As a result, their actual effectiveness remains unclear.
    To address this issue, we propose an enhanced implementation of Hottopixx
    designed to reduce computational time and improve endmember extraction performance.
    We demonstrate its  effectiveness through experiments.
    The results suggest that our implementation enables the application of Hottopixx
    for	endmember extraction from real hyperspectral images and
    allows us to achieve reasonably high accuracy in estimating endmember signatures.
\end{abstract}


\section{Introduction} \label{Sec: introduction}
A hyperspectral camera is an optical instrument used to measure the spectra of materials in a scene.
The resulting hyperspectral images (HSIs) typically consist of several hundred spectral bands, enabling detailed analysis of material composition.
Endmember extraction from HSIs aims to identify the spectral signatures of materials (i.e., the major components in observed scenes).
This process is a key step in utilizing hyperspectral imaging for practical applications.
For a comprehensive overview of the latest advancements in hyperspectral image analysis, refer to the tutorial paper \cite{Gha18}.

Self-dictionary methods are known as promising approaches for extracting endmembers from HSIs.
These methods formulate the task as a sparse optimization problem using the data itself as a dictionary,
and then employ convex relaxation techniques to solve it.
In 2012, Bittorf et al.\ \cite{Bit12} developed the Hottopixx method, originally designed for extracting hot topics from documents.
This method can be viewed as a self-dictionary approach for the task of extracting endmembers from HSIs.
Since the work of Bittorf et al., several authors \cite{Gil13, Gil14b, Miz22} have refined Hottopixx
and examined the performance of their refinements both theoretically and practically.

Hottopixx methods employ a linear programming (LP) problem, referred to as the Hottopixx model,
as a relaxation of the sparse optimization formulation.
Typically, these methods involve two main steps: first, compute the optimal solution to the Hottopixx model;
second, identify endmembers based on information obtained from the optimal solution.
The second step serves as a postprocessing procedure for Hottopixx.
In \cite{Gil13,  Miz22}, the authors proposed a clustering algorithm for postprocessing,
which improves the endmember extraction performance of Hottopixx methods.
They provided theoretical evidence to support this approach in their study.

However, the practical application of Hottopixx methods is hindered
by high computational costs associated  with solving Hottopixx models,
as the size of the models grows quadratically with the number of pixels in the HSI.
As a result, it is unclear whether these methods are actually effective in extracting endmembers from HSIs.
To address this issue, we propose an enhanced implementation of the Hottopixx method of \cite{Miz22}.
As summarized below, our approach can significantly reduce computational time and improve the accuracy of the estimated endmember signatures.
Therefore, we refer to it as EEHT, an efficient and effective implementation of Hottopixx for the endmember extraction task.

The contributions of this study are summarized as follows.
\paragraph{Reducing the computational time of Hottopixx}
Solving Hottopixx models is computationally costly, which poses an obstacle to applying them to the endmember extraction task.
To remedy this issue, we propose a row and column expansion (RCE) algorithm for solving Hottopixx models efficiently.
This algorithm follows the framework of column generation,
a classical but powerful technique for solving large-scale LPs.

Many zero elements may exist in the optimal solution of a Hottopixx model.
This suggests that Hottopixx models can be solved by breaking them up into smaller subproblems and solving them.
We present Theorem \ref{Thm: main}, which shows that
the optimal solution of the full Hottopixx model can be obtained by solving its subproblems.
We then develop an RCE algorithm based on this theoretical result.
The details are described in Section \ref{Sec: solving Hottopixx models efficiently}.

\paragraph{Enhancing the endmember extraction performance of Hottopixx}
The postprocessing step in the Hottopixx method described in \cite{Miz22} conducts data clustering and outputs one element from each cluster.
The choice of elements affects the endmember extraction performance of Hottopixx.
The methods used in the previous studies \cite{Gil13, Gil14b, Miz22} were based on the optimal solutions to Hottopixx models.
Alternatively, we propose a method based on the shape of clusters;
it computes the centroids of each cluster and chooses elements close to them.
This method is simple but effective in enhancing the endmember extraction performance of Hottopixx.
We will call it the cluster centroid choice.
The details are described in Section \ref{Sec: detailed description of EEHT}.

\paragraph{Demonstrating the performance of EEHT by experiment}
We developed EEHT by incorporating RCE and the cluster centroid choice.
We then experimentally tested its performance on endmember extraction problems.
The first experiment (Section \ref{Subsec: computational efficiency}) examined the computational time of RCE.
The second experiment (Section \ref{Subsec: endmember extraction performance})
evaluated the endmember extraction performance of EEHT on synthetic HSI datasets.
These datasets were constructed from two real HSIs, Jasper Ridge and Samson.
We compared EEHT with six existing methods, including MERIT from \cite{Ngu22},
which is a method based on a self-dictionary, similar to EEHT.
Finally, we conducted an experimental study (Section \ref{Subsec: unmixing of Urban})
on hyperspectral unmixing of the Urban HSI dataset.
The results show that, compared with the existing methods,
EEHT usually provides more accurate estimates of endmember signatures for Urban.

\subsection{Organization of This Paper}
This paper is organized as follows.
Section \ref{Sec: related work} reviews related work on endmember extraction methods.
Section \ref{Sec: preliminaries} introduces the preliminaries.
Section \ref{Sec: problem and methods} formulates the endmember extraction problem and explains the algorithm of Hottopixx.
Section \ref{Sec: solving Hottopixx models efficiently} and Section \ref{Sec: detailed description of EEHT}
present the details of the RCE and EEHT.
Section \ref{Sec: experiments} describes the experiments.

\section{Related Work} \label{Sec: related work}
Numerous methods have been proposed for extracting endmembers from HSIs,
as detailed in the survey papers \cite{Bio12, Ma14} and the textbook \cite{Gil20}.
These methods can be categorized into two groups: convex optimization-based methods and greedy methods.
Hereinafter, we use the term ``convex method" as a shorthand to refer to the convex optimization-based method.
Convex methods are robust to noise but computationally expensive, whereas greedy methods are faster but less robust.
Hottopixx is an example of a convex method.

\subsubsection{Convex Methods}
As discussed in Section \ref{Subsec: Hottopixx methods},
the endmember extraction problem can be formulated as a sparse optimization problem.
However, directly solving sparse optimization problems is computationally intractable.
Thus, we instead solve their convex relaxation problems.
A typical convex method first computes the optimal solution to the relaxed problem
and then identifies the endmembers based on this solution.
Hottopixx is closely related to FGNSR, proposed by Gillis and Luce \cite{Gil18},
and MERIT, proposed by Nguyen, Fu and Wu \cite{Ngu22}, both of which are convex methods using self-dictionaries.
We will review FGNSR and MERIT in more detail in Section \ref{Subsec: convex methods in prior work}.

\subsubsection{Greedy Methods}
Greedy methods choose the columns of the HSI matrix one by one on the basis of certain criteria.
One of the most popular greedy methods for endmember extraction is SPA,
which dates back to the work of Ara\'{u}jo et al.\ \cite{Ara01} in 2001.
Subsequently, Gillis and Vavasis \cite{Gil14a} gave a theoretical justification for the robustness of SPA to noise.
SPA recursively chooses columns of the input matrix.
For a column chosen in the previous iteration,
it projects columns onto the orthogonal complement of the chosen column
and then chooses one with the maximum $L_2$ norm.
To enhance the robustness of SPA to noise,
Gillis and Vavasis \cite{Gil15} developed PSPA, which is SPA with preconditioning, and
Mizutani \cite{Miz14} ER, which is SPA with a preprocessing.
Both preconditioning and preprocessing use singular value decomposition (SVD) based dimensionality reduction and minimum volume ellipsoids.
Gillis \cite{Gil14c} proposed SNPA, which is similar to SPA but it uses a projection
onto the convex hull of the origin and previously chosen columns.
Nascimento and Bioucas-Dias \cite{Nas05} proposed VCA, which
applies an SVD-based dimensionality reduction to the input matrix, and then,
runs the same algorithm as SPA except for the criterion for choosing columns.
Other popular methods in the remote sensing community include
PPI \cite{Boa95}, N-FINDR \cite{Win99}, and AVMAX and SVMAX \cite{Cha11}.

\section{Preliminaries} \label{Sec: preliminaries}
\subsection{Notation} \label{Subsec: notation}
$\zero$ denotes a vector of all zeros,
$\one$ a vector of all ones,
$O$ a matrix of all zeros,
$I$ the identity matrix,
and $J$ a matrix of all ones.
$\Real^{d \times n}_+$ represents the set of nonnegative matrices.

Let $\a \in \Real^d$ and $A \in \Real^{d \times n}$.
The notation $\a(i)$ indicates the $i$th entry of $\a$.
Similarly, $A(i,j)$ indicates the $(i,j)$th entry of $A$,
$A(i, :)$ the $i$th row, and $A(:,j)$ the $j$th column.
The $j$th column is also denoted by $\a_j$.
We write $\trace(A)$ for the trace of $A$.
The notation $\diag(\a)$ refers to a diagonal matrix of size $d$ containing the elements of $\a$ in the diagonal positions.
Let $A$ be a square matrix, i.e., $d = n$.
In this case, $\diag(A)$ refers to a vector of size $d$ consisting of the diagonal elements of $A$.

We use $\NC$ to denote $\{1, \ldots, n\}$.
Let $A = [\a_1, \ldots, \a_n] \in \Real^{d \times n}$.
For $\IC \subset \NC$,
we use $A(\IC)$ or $[\a_i \mid i \in \IC]$ to denote the submatrix of $A$ indexed by $\IC$ in ascending order.
Specifically, if $\IC = \{i_1, \ldots, i_p\}$ with $i_1 < \cdots < i_p$, then
$A(\IC) = [\a_i \mid i \in \IC] = [\a_{i_1}, \ldots, \a_{i_p}]$.

We use $\| \cdot \|$ to denote any norm of a vector or a matrix.
In particular,
$\| \cdot \|_p$ refers to the $L_p$-norm of a vector or a matrix, and $\| \cdot \|_F$ the Frobenius norm of a matrix.
For $A \in \Real^{d \times n}$, we define $\| A \|_{p,q} = \sum_{i=1}^{n} \| A(i, :) \|_p^q$ and
use $\| A \|_{\mathrm{row}, 0}$ to denote the number of nonzero rows in $A$.
Moreover, $\langle A, B \rangle$ indicates the inner product of two matrices $A$ and $B$ of the same size,
defined by $\langle A, B \rangle = \trace(A^\trans B)$.

For $A \in \Real^{d \times n}$ (resp. $\a \in \Real^d$), we define $A^+$ (resp. $\a^+$) to be the matrix (resp. vector)
obtained by replacing all negative values of $A$ (resp. $\a$) with zero.
Moreover, we define $A^-$ by $A^- = A^+ - A$. Hence, $A^+$ and $A^-$ satisfy $A^+, A^- \ge O$ and $A = A^+ - A^-$.
The $L_1$ norm of $A$ can be written using $A^+$ and $A^-$ as follows:
\begin{align} \label{Eq: L1 norm expression}
    \| A \|_1 = \max_{j=1, \ldots, n} \sum_{i=1}^{d} A^+(i,j) + A^-(i,j).
\end{align}

Additionally, for an optimization problems $\OptProb$,
we denote the optimal value of $\OptProb$ by $\opt(\OptProb)$.
For a set $\IC$, we use $|\IC|$ to denote the number of elements in $\IC$.

\subsection{Duality and Solution Methods for LP Problems} \label{Subsec: duality and solution methods for LPs}
As mentioned in Section \ref{Sec: introduction},
we propose the RCE algorithm for solving Hottopixx models efficiently.
This algorithm is based on the duality of LP problems.
For $A \in \Real^{m \times n}$, $\b \in \Real^m$, and $\c \in \Real^n$,
consider an LP problem in the standard form and its corresponding dual problem, given as follows:
\begin{alignat*}{5}
    \text{(Primal)} & \quad & \min_{\x \in \Real^n} & \quad & \c^\trans \x & \quad & \text{s.t.} & \quad & A\x = \b, \ \x \ge \zero, \\
    \text{(Dual)}   &       & \max_{\y \in \Real^m} & \quad & \b^\trans \y & \quad & \text{s.t.} & \quad & A^\trans \y \le \c.
\end{alignat*}
\begin{itemize}
    \item \textit{The duality theorem:}
          If the primal and dual are feasible,
          then there exist optimal solutions $\x$ and $\y$ to the primal and dual and
          they satisfy $\c^\trans \x = \b^\trans \y$.
    \item \textit{Implication of the weak duality theorem:}
          Let $\x$ and $\y$ be feasible solutions to the primal and dual.
          If $\c^\trans \x = \b^\trans \y$,
          then $\x$ and $\y$ are optimal solutions to the primal and dual.
\end{itemize}
For more details, see the textbooks \cite{Ber97, Ben01}.
We will use these results to develop RCE.

In solving LPs, the simplex method and the interior-point method are mainly used.
The iteration complexity has been established for the interior-point method.
For instance, the primal-dual interior-point method computes the optimal solutions of
the primal and dual problems simultaneously,
where the iteration count is $O(\sqrt{n} \log(1/\epsilon))$
for a given tolerance $\epsilon > 0$,
and each iteration involves solving a system of linear equations with $2n + m$ variables and $2n + m$ equations;
see the textbook \cite{Wri97} for further details.

\section{Self-Dictionary Approaches for Solving Endmember Extraction Problems} \label{Sec: problem and methods}
\subsection{Endmember Extraction of HSIs} \label{Subsec: endmember extracion of HSIs}
Suppose that we are given HSI data for some area consisting of $n$ pixels
acquired by a hyperspectral camera with $d$ spectral bands.
We represent the HSI as a matrix $A \in \Real^{d \times n}$ where
the $(i,j)$th entry of $A$ stores the measurement at the $i$th band and $j$th pixel.
In this paper, we refer to such a matrix $A$ an {\it HSI matrix}.
The $j$th column $\a_j$ of $A$ corresponds to the observed spectral signature at the $j$th pixel.

Let $A = [\a_1, \ldots, \a_n] \in \Real^{d \times n}$ be an HSI matrix,
and consider a linear mixing model (LMM) for $A$.
Under the LMM, the observed spectral signatures $\a_1, \ldots, \a_n$ at each pixel are written as
\begin{align*}
    \a_j = \sum_{i=1}^{r} h_{ij} \w_i + \v_j, \ j = 1, \ldots, n.
\end{align*}
Here, $\w_i$ satisfies $\w_i \ge \zero$ and represents the $i$th endmember signature;
$h_{ij}$ satisfies $\sum_{i=1}^{r} h_{ij} = 1, \ h_{ij} \ge 0$ and represents the abundance fraction of
the $i$th endmember signature at the $j$th pixel; $\v_j$ represents noise.
The equation above can be expressed in matrix form as
\begin{align} \label{Eq: LMM}
    A = W H + V
\end{align}
by letting $W = [\w_1, \ldots, \w_r] \in \Real^{d \times r}_+$,
$H \in \Real^{r \times n}_+$ such that $H(i,j) = h_{ij}$, and $V = [\v_1, \ldots, \v_n] \in \Real^{d \times n}$.
We call $W$ and $H$ the {\it endmember matrix} and the {\it abundance matrix} of $A$, respectively.

\textit{Endmember signatures}, represented by $\w_1, \ldots, \w_r$, are the spectral signatures of certain materials
contained in the image scene.
The term \textit{endmember} is used for referring to the corresponding material;
hence, endmembers are materials contained in the image scene and represent the major components of it.

We will make some assumptions about the HSIs.
We say that a pixel is \textit{pure} if it contains a single material corresponding to some endmember.
We assume that there is at least one pure pixel for every endmember.
This is called the \textit{pure pixel assumption}.
Accordingly, the HSI matrix $A$ shown in (\ref{Eq: LMM}) can be rewritten as
\begin{align} \label{Eq: LMM under PPA}
    A = W[I, \bar{H}] \Pi + V
\end{align}
for $W \in \Real^{d \times r}_+, \bar{H} \in \Real^{r \times (n-r)}_+, V \in \Real^{d \times n}$
and a permutation matrix $\Pi$ of size $n$.
If $A$ is noiseless, i.e., $V = O$, then, we have $A = W[I, \bar{H}] \Pi$, which implies that
all columns of $W$ appear among those of $A$.
Furthermore, we assume that the number of endmembers in an HSI is known in advance.
We can now formulate the endmember extraction problem as follows.

\begin{framed}
    \begin{prob} \label{Prob: endmember extraction}
        Given the HSI matrix $A$ shown in (\ref{Eq: LMM under PPA}) and the number $r$ of endmembers,
        find $r$ columns of $A$ close to those of the endmember matrix $W$.
    \end{prob}
\end{framed}

The pure pixel assumption is considered reasonable.
Meanwhile, it is difficult to determine the exact number of endmembers in an HSI.
Thus, in our experiments, we use the estimate of the number of endmembers provided by previous studies.
Problem \ref{Prob: endmember extraction} is known to be equivalent to
the problem of computing a nonnegative matrix factorization (NMF) under the separability assumption.
Arora et al.\ \cite{Aro12a} studied the separable NMF problems from a theoretical point of view.
Their results tell us that we can find $W$ from $A$ exactly if $A$ is noiseless;
otherwise, if $A$ contains noise $V$ so that the amount of $V$ is smaller than some level,
we can find $r$ columns of $A$ close to those of $W$.
Given $A$ as shown in (\ref{Eq: LMM under PPA}),
we say that an algorithm for Problem \ref{Prob: endmember extraction} is {\it robust to noise}
if it can find $r$ such columns of $A$.
For further reading on separable NMF, refer to \cite{Fu19, Gil20}.

Once Problem \ref{Prob: endmember extraction} is solved, we can compute the abundance matrix $H$ of $A$
from the estimated endmember signatures. Let $\IC$ be the index set of $r$ columns found in the problem.
Then, $H$ is the optimal solution to the following convex optimization problem:
\begin{alignat}{4} \label{Prob: abundance computation}
     & \min_{X \in \Real^{r \times n}} & \quad & \| A - A(\IC) X \|_F^2 & \quad & \text{s.t.} & \quad & \one^\trans X = \one^\trans, \ X \ge O.
\end{alignat}

We need to evaluate the accuracy of the estimated endmember signatures.
In this paper, we use the mean-removed spectral angle (MRSA) of two spectral signatures for this evaluation.
For $\c \in \Real^d$, we set $\ave(\c) = (\one^\trans \c / d) \cdot \one \in \Real^d$.
For spectral signature vectors $\a, \b \in \Real^d$,
the MRSA is defined as
\begin{align*}
    \MRSA(\a, \b) =
    \frac{1}{\pi} \arccos \frac{(\a - \ave(\a))^\trans (\b -  \ave(\b))}{\| \a - \ave(\a) \|_2 \| \b - \ave(\b) \|_2},
\end{align*}
which takes values in the interval from 0 to 1.
A smaller MRSA value for $\a$ and $\b$ means that $\a$ is more similar to $\b$.

\subsection{Hottopixx Methods} \label{Subsec: Hottopixx methods}
Dictionary learning methods are often used for extracting endmembers from HSIs.
Let us consider the input matrix $A$ of Problem \ref{Prob: endmember extraction} as a dictionary
and the columns of $A$ as the atoms of the dictionary.
The problem can thus be restated as one of choosing $r$ atoms from the dictionary
to achieve a good approximation of the dictionary by means of nonnegative linear combinations of the chosen atoms.
Since the input matrix $A$ itself is used as a dictionary,
it is referred to as a \textit{self-dictionary}.
It should be noted that the pure pixels assumption is made for HSIs in Problem \ref{Prob: endmember extraction};
thus we can use $A$ as a dictionary. In cases where this assumption does not hold, see Remark \ref{Rem: construction of dictionary}.

From the perspective of dictionary learning,
let us formulate Problem \ref{Prob: endmember extraction} as a sparse optimization problem,
\begin{subequations} \label{Prob: sparse optimization formulation}
    \begin{alignat}{4}
         & \min_{X \in \Real^{n \times n}} & \quad & \| A - A X \|                                                             \\
         & \quad \text{s.t.}               & \quad & \| X \|_{\mathrm{row}, 0} = r, \ \one^\trans X = \one^\trans,  \ X \ge O.
    \end{alignat}
\end{subequations}
This problem is difficult to solve since it includes the combinatorial constraint $\| X \|_{\mathrm{row}, 0} = r$.
Hottopixx methods were initially proposed by Bittorf et al.\ \cite{Bit12} in 2012 in the context of topic modeling.
Subsequently, several refinements \cite{Gil13, Gil14b, Miz22} have been developed.
These methods are based on the LP relaxation of problem (\ref{Prob: sparse optimization formulation}).

In this paper, we propose an implementation of the Hottopixx method that was proposed in \cite{Miz22}.
To this end, we review the details of the Hottopixx method below.
For a given $A$ and $r$ in Problem \ref{Prob: endmember extraction},
the first step of the Hottopixx method involves constructing an optimization problem,
\begin{alignat*}{5}
    \HT: & \quad & \min_{X \in \Real^{n \times n}} & \quad & \| A - AX \|_1 & \quad & \text{s.t.} & \quad & X \in \FC.
\end{alignat*}
The feasible region $\FC$ is defined by
\begin{alignat*}{2}
     & \sum_{i=1}^{n} X(i,i) = r,     & \quad &                     \\
     & 0 \le X(i,j) \le X(i,i) \le 1, &       & i,j = 1, \ldots, n.
\end{alignat*}
In this paper, we call $\HT$ the \textit{Hottopixx model},
although it differs from the original Hottopixx model proposed by Bittorf et al.\ \cite{Bit12}.
As shown in Section \ref{Subsec: subproblems of H}, $\HT$ can be reduced to an LP problem.
The previous study \cite{Miz22} directly solved this LP using a CPLEX solver.
The constraints of $\HT$ relax the constraint $\| X \|_{\mathrm{row}, 0} = r$ of problem (\ref{Prob: sparse optimization formulation}).
However, it is possible for many rows of the optimal solution $X$ in $\HT$ to be zero, since
the constraints of $\HT$ imply that
the $i$th row of $X$ is zero if $X(i,i) = 0$ and many of $X(1,1), \ldots, X(n,n)$ could be zero.

Problem (\ref{Prob: sparse optimization formulation}) includes the sum-to-one constraint $\one^\trans X = \one^\trans$,
but we exclude it from $\HT$, as the original Hottopixx model does not include it.
As described in Remark 1 of \cite{Gil13},
theoretical studies suggest that
the endmember extraction performance of Hottopixx methods can be further enhanced by incorporating the sum-to-one constraint into $\HT$.

The second step involves choosing $r$ columns of $A$ using the optimal solution $X$ of $\HT$.
A simple way of doing so is to find a set $\IC$ of $r$ indices corresponding to the $r$ largest elements of $\diag(X)$ and,
then, output columns $\a_i$ with $i \in \IC$.
Theorem 3.1 of \cite{Miz22} ensures that the algorithm is robust to noise.
However, the theorem is invalid and does not hold in a case where there are overlaps of pure pixels in the HSIs.
To cope with this issue, we use a clustering technique.
Algorithm \ref{Alg: HT} describes each step of Hottopixx with postprocessing \cite{Miz22}.
A cluster $\SC_\ell$ in the algorithm refers to a subset of the column index set $\NC$ of $A$.
Due to page limitations,
a detailed description of how to construct clusters is provided in Appendix \ref{Appx: Cluster construction in step 2-1 of Algorithm 1}.

\begin{algorithm}[h]
    \caption{Hottopixx with postprocessing (Algorithm 5.1 of \cite{Miz22})}
    \label{Alg: HT}
    \smallskip
    Input: $A = [\a_1, \ldots, \a_n] \in \Real^{d \times n}$ and a positive integer $r$. \\
    Output: $r$ columns of $A$.
    \medskip
    \begin{enumerate}[1:]
        \item Compute the optimal solution $X$ of $\HT$.
        \item Set $\p_1 = \diag(X), \IC = \emptyset$ and $\ell = 1$. Perform the following procedure.
              \begin{enumerate}[{2-}1:]
                  \item Construct a cluster $\SC_\ell$ using $\p_\ell$; see Appendix \ref{Appx: Cluster construction in step 2-1 of Algorithm 1} for the details.
                  \item Choose one element from $\SC_\ell$ and add it to $\IC$. Increase $\ell$ by $1$.
                  \item If $\ell = r$, return $\a_i$ with $i \in \IC$ and terminate;
                        otherwise, construct $\p_\ell$ as
                        \begin{align*}
                            \p_\ell(u) =
                            \left\{
                            \begin{array}{ll}
                                0       & \text{if} \ u \in \SC_1 \cup \cdots \cup \SC_{\ell-1}, \\
                                \p_1(u) & \text{otherwise,}
                            \end{array}
                            \right.
                        \end{align*}
                        and go to step 2-1.
              \end{enumerate}
    \end{enumerate}
\end{algorithm}
Algorithm \ref{Alg: HT} has an advantage over other algorithms \cite{Bit12, Gil13, Gil14b}
in that it does not require us to specify the noise level involved in $A$ as input.
Theorem 3.2 of \cite{Miz22} ensures that Algorithm \ref{Alg: HT} is robust to noise.
Even if there are overlaps among pure pixels in the HSI, the theorem remains valid and the algorithm is robust to noise.
In most cases, there are a number of pixels in the HSIs that are close to pure pixels.
Specifically, we will see this feature in the Urban dataset in Section \ref{Subsec: unmixing of Urban}.
Hence, when solving Problem \ref{Prob: endmember extraction}, we should use the postprocessing procedure described in step 2 of Algorithm \ref{Alg: HT}.

\begin{remark} \label{Rem: construction of dictionary}
    If the pure pixel assumption does not hold for HSIs, the input matrix $A$ of Problem \ref{Prob: endmember extraction} no longer serves as a dictionary.
    In such cases, since there are libraries containing numerous spectral signatures of materials,
    as proposed in \cite{Ior11, Ior14a, Ior14b} it is feasible to construct a dictionary by collecting samples from these libraries.
\end{remark}

\subsection{Convex Methods in Prior Work} \label{Subsec: convex methods in prior work}
We review convex methods from prior work,
focusing in particular on FGNSR \cite{Gil18}, and MERIT \cite{Ngu22}.
To begin, we recall problem (\ref{Prob: sparse optimization formulation}),
which is the sparse optimization formulation for Problem \ref{Prob: endmember extraction}.
FGNSR and MERIT solve its convex relaxation as follows:
\begin{alignat}{4} \label{Prob: self-dictionary sparse regression}
    \min_{X \in \Real^{n \times n}} & \quad & \frac{1}{2} \| A - AX \|_F^2 + \lambda \cdot \Phi(X) & \quad & \mbox{s.t.} & \quad & X \in \GC
\end{alignat}
for a penalty parameter $\lambda \ge 0$,
where $\Phi$ is the regularization term and $\GC$ is the feasible region.
Here, $\Phi$ is designed to promote row sparsity in $X$, that is, to minimize the number of nonzero rows.

FGNSR uses $\Phi(X) = \taub^\trans \diag(X)$, where $\taub \in \Real^n$ is close to $\one$.
It uses $\GC$ defined by
\begin{alignat*}{2}
     & X(i,i) \le 1,              & \quad & i  = 1, \ldots, n,  \\
     & c_i X(i,j) \le c_j X(i,i), &       & i,j = 1, \ldots, n, \\
     & X \ge O                    &       &
\end{alignat*}
where $c_i = \| A(:,i) \|_1$.
Problem (\ref{Prob: self-dictionary sparse regression}) with $\Phi$ and $\GC$ as described above
is closely related to the original Hottopixx model by Bittorf et al.\ \cite{Bit12}.
Gillis and Luce \cite{Gil18} showed that the projection of a matrix onto $\GC$ can be computed in $O(n^2 \log n)$.
Taking this into account it, they developed a first-order method equipped with Nesterov's acceleration to solve these problems.

MERIT uses $\Phi(X) = \sum_{i=1}^{n} \phi_{\mu} (X(i,:))$, where
\begin{align}
    \phi_{\mu}(\x) = \mu \log \left( \frac{1}{n} \sum_{i=1}^{n} \exp \frac{x_i}{\mu} \right) \label{Eq: phi}
\end{align}
for $\x = [x_1, \ldots, x_n]^\trans \in \Real^n$ and a parameter $\mu > 0$.
It uses $\GC$ defined by $\one^\trans X = \one^\trans$ and $X \ge O$, referred to as the simplex constraint.
The function $\Phi(X)$ serves as a smooth approximation of $\| X \|_{\infty, 1}$, as the relation
$\| \x \|_{\infty} - \mu \log n \le \phi_{\mu}(\x) \le \| \x \|_{\infty}$ holds.
Accordingly, minimizing $\Phi(X)$ causes many rows of $X$ to be zero when $\mu$ is small.
Nguyen, Fu and Wu \cite{Ngu22} proposed to solve such problems with the Frank-Wolfe method.
They also examined the theoretical performance of MERIT in their study.

Convex methods for solving Problem \ref{Prob: endmember extraction} and related problems
can be found in the works of Elhamifar, Sapiro and Vidal \cite{Elh12}, and of Esser et al.\ \cite{Ess12}.
Fu and Ma \cite{Fu16b} analyzed the performance of convex method based on problem (\ref{Prob: self-dictionary sparse regression})
with $\Phi(X) = \| X \|_{\infty, q}$ for $0 < q \le 1$ and with $\GC$ defined by the simplex constraint.
Fu, Sidiropoulos and Ma \cite{Fu16a} applied convex optimization approaches to power spectral separation.

\section{Solving Hottopixx Models Efficiently} \label{Sec: solving Hottopixx models efficiently}
\subsection{Algorithm Outline} \label{Subsec: algorithm outline}
In Section \ref{Subsec: subproblems of H}, we show that $\HT$ can be reduced to an LP with $O(n^2)$ variables and $O(n^2)$ constraints
for HSI matrices $A \in \Real^{d \times n}$ with $d \le n$.
Accordingly, when $n$ is large, directly solving the LP, as in the previous study \cite{Miz22}, becomes  computationally challenging.
To remedy this computational issue, we develop the RCE algorithm for solving $\HT$ efficiently
based on the framework of column generation, which is known as a classical but powerful technique for solving large-scale LPs;
see, for instance, the textbook \cite{Ber97} for details on column generation methods.
The algorithm is based on the observation that there may be many zero rows in the optimal solution of $\HT$.
We can significantly reduce the size of $\HT$ by exploiting the sparsity of the optimal solution.

Let $\LC$ and $\MC$ be the subsets of $\NC$, where recall that $\NC$ is defined as $\{1, \ldots, n\}$ in Section \ref{Subsec: notation}.
Hereinafter, we use $\ell$ and $m$ to denote $|\LC|$ and $|\MC|$, respectively.
RCE uses a subproblem of $\HT$ with a variable $X \in \Real^{\ell \times m}$.
The feasible region $\FC(\ell, m)$ is defined by
\begin{alignat*}{2}
     & \sum_{i=1}^{\ell} X(i,i) = r,  & \quad &                                          \\
     & 0 \le X(i,j) \le X(i,i) \le 1, &       & i = 1, \ldots, \ell, \ j = 1, \ldots, m.
\end{alignat*}
We consider the following subproblem:
\begin{alignat*}{5}
     & \HT' : & \quad       & \min_{X \in \Real^{\ell \times \ell}} & \quad                  & \| A(\LC) - A(\LC)X \|_1
     & \quad  & \text{s.t.} & \quad                                 & X \in \FC(\ell, \ell).
\end{alignat*}
$\HT'$ has fewer variables and constraints than the original problem $\HT$.
Hence, solving $\HT'$ is computationally cheaper than solving $\HT$.
We show in Theorem \ref{Thm: main} the conditions under which
the optimal solution of $\HT'$ yields the optimal solution of $\HT$.
Based on this theorem, we design an algorithm for solving $\HT$ efficiently, which can be outlined as follows:
solve $\HT'$ and verify the conditions; if the conditions are not satisfied,
update $\LC$ by adding some elements in $\NC \setminus \LC$ to $\LC$ and solve $H'$ again.
We call this algorithm ``row and column expansion'', RCE,
since it starts from the optimal solution $X$ of $\HT'$, and then expands the rows and the columns of $X$ step-by-step.

\subsection{Subproblem of $\HT$} \label{Subsec: subproblems of H}
Here, we formally describe the subproblem of $\HT$.
For given subsets $\LC$ and $\MC$ of $\NC$ satisfying $\LC \subset \MC$,
choose a permutation matrix $\Pi \in \Real^{m \times m}$ such that
\begin{align*}
    A(\MC)\Pi = \ALM.
\end{align*}
Then, construct the subproblem
\begin{alignat*}{3}
     & \HT(\LC, \MC) : & \quad & \min_{X \in \Real^{\ell \times m}} & \quad & \| A(\MC)\Pi - A(\LC)X \|_1 \\
     &                 & \quad & \quad \text{s.t.}                  & \quad & X \in \FC(\ell, m).
\end{alignat*}
The permutation matrix $\Pi$ is chosen to be the identity matrix when $\LC = \MC$.
Hence, $\HT$ coincides with $\HT(\NC, \NC)$.
We can reduce $\HT(\LC, \MC)$ to the following LP problem,
\begin{alignat*}{4}
     & \Primal(\LC, \MC) : & \quad & \text{min}  & \quad & u                                                               \\
     &                     &       & \text{s.t.} &       & A(\MC)\Pi - A(\LC)X = F - G,                                    \\
     &                     &       &             &       & \sum_{i=1}^{d} F(i,j) + G(i,j) \le u,  \quad  j = 1, \ldots, m, \\
     &                     &       &             &       & F \ge O, \ G \ge O, \ X \in \FC(\ell, m).
\end{alignat*}
Here, $(X, F, G, u) \in \Real^{\ell \times m} \times \Real^{d \times m} \times \Real^{d \times m} \times \Real$ is the variable.
In this problem, the number of variables is $\ell m + 2dm +1$,
the number of equality constraints is $dm + 1$,
and the number of inequality constraints is $\ell m + \ell + m$.
By introducing slack variables to the inequality constraints, we convert $\Primal(\LC, \MC)$ to the standard form LP.
The corresponding dual problem is given by:
\begin{alignat*}{4}
     & \Dual(\LC, \MC) : & \quad & \text{max}  & \quad & \langle A(\MC)\Pi, Y \rangle + r  v - \one^\trans \t \\
     &                   &       & \text{s.t.} &       & A(\LC)^\trans Y + [v I, O] -  [\diag(\t), O]         \\
     &                   &       &             &       & \qquad - Z^\trans + [\diag(Z^\trans \one), O] \le O, \\
     &                   &       &             &       & - J  \cdot \diag(\s) \le Y \le  J  \cdot \diag(\s),  \\
     &                   &       &             &       & \one^\trans \s \le 1,                                \\
     &                   &       &             &       & \s \ge \zero, \ \t \ge \zero, \ Z \ge O.
\end{alignat*}
Here,  $(Y, Z, \s, \t, v) \in \Real^{d \times m} \times \Real^{m \times \ell} \times \Real^m \times \Real^\ell \times \Real$ is the variable.
In this problem, the number of variables is $\ell m + dm + \ell + m + 1$,
and the number of inequality constraints is $\ell m + 2dm + 1$.

To prove Theorem \ref{Thm: main},
it is necessary to establish a connection among $\Primal(\LC, \MC)$, $\Dual(\LC, \MC)$, and $\HT(\LC, \MC)$.
This relationship is formalized in the following theorem.
\begin{thm} \label{Thm: optimal values of S, P, D}
    For any subsets $\LC$ and $\MC$ of $\NC$ satisfying $r \le \ell \le m$, the following hold.
    \begin{enumerate}[(i)]
        \item $\opt(\Primal(\LC, \MC)) = \opt(\Dual(\LC, \MC))$.
        \item $\opt(\Primal(\LC, \MC)) = \opt(\HT(\LC, \MC))$.
        \item If $(X^*, F^*, G^*, u^*)$ is the optimal solution to $\Primal(\LC, \MC)$,
              then $X^*$ is the optimal solution to $\HT(\LC, \MC)$.
        \item If $X^*$ is the optimal solution to $\HT(\LC, \MC)$,
              then, $(X^*, R^+, R^-, \|R\|_1)$ for $R = A(\MC)\Pi - A(\LC)X^*$ is the optimal solution to $\Primal(\LC, \MC)$.
    \end{enumerate}
\end{thm}
The proof is provided in Appendix \ref{Appx: Proof of Thm: optimal values of S, P, D}.

\subsection{Description of the RCE Algorithm} \label{Subsec: description of RCE}
The optimal solution to $\HT$ can be obtained by solving $\Primal(\LC,\LC)$ and $\Dual(\LC, \LC)$ if certain conditions are satisfied.
Let $\alpha^*$ be the optimal solution to $\Primal(\LC,\LC)$, and let $\beta^*$ be the optimal solution to $\Dual(\LC,\LC)$.
If $\alpha$ and $\beta$ can be constructed from $\alpha^*$ and $\beta^*$, respectively, such that
\begin{itemize}
    \item $\alpha$ is a feasible solution to $\Primal(\NC, \NC)$,
    \item $\beta$ is a feasible solution to $\Dual(\NC, \NC)$, and
    \item the objective function values attained by $\alpha$ and $\beta$ coincide,
\end{itemize}
then, by the weak duality theorem reviewed in Section \ref{Subsec: duality and solution methods for LPs},
$\alpha$ is the optimal solution to $\Primal(\NC, \NC)$.
From Theorem \ref{Thm: optimal values of S, P, D}, we observe that the optimal solution to $\HT$ can be obtained from $\alpha$.
We formalize this claim in the following theorem.
To do so, we introduce an auxiliary problem:
for a given matrix $X^* \in \Real^{\ell \times \ell}$, construct
\begin{alignat*}{5}
     & \AugProb_j(\LC, X^*) : & \quad & \min_{\gammab_j \in \Real^\ell} & \quad & \| \a_j - A(\LC) \gammab_j \|_1    \\
     &                        &       & \ \text{s.t.}                   & \quad & \zero \le \gammab_j \le \diag(X^*)
\end{alignat*}
for $j \in \NC \setminus \LC$.

\begin{thm} \label{Thm: main}
    Let $\LC \subset \NC$.
    Let $\alpha^* = (X^*, F^*, G^*, u^*)$ be the optimal solution to $\Primal(\LC,\LC)$ and
    $\beta^* = (Y^*, Z^*, \s^*, \t^*, v^*)$	be the optimal solution to $\Dual(\LC,\LC)$.
    Consider the following conditions regarding $\alpha^*$ and $\beta^*$.
    \begin{enumerate}[(C1)]
        \item $X^*$ of $\alpha^*$ satisfies
              \begin{align*}
                  \opt(\AugProb_j(\LC, X^*)) \le \opt(\Primal(\LC, \LC))
              \end{align*}
              for every $j \in \NC \setminus \LC$.
        \item $Y^*$ and $v^*$ of $\beta^*$ satisfy
              \begin{align*}
                  v^* + \one^\trans ((Y^*)^\trans \a_j)^+ \le 0
              \end{align*}
              for every $j \in \NC \setminus \LC$.
    \end{enumerate}
    The following hold.
    \begin{enumerate}[(i)]
        \item Assume that condition (C1) holds. Let $\Gamma^* = [\gammab^*_j  \mid j \in \NC \setminus \LC]$
              for the optimal solution $\gammab^*_j$ to $\AugProb_j(\LC, X^*)$. Then,
              \begin{itemize}
                  \item $\opt(\HT(\LC, \LC))  = \opt(\HT(\LC, \NC))$, and
                  \item the matrix $[X^*, \Gamma^*]$ is the optimal solution to $\HT(\LC, \NC)$.
              \end{itemize}
        \item Assume that conditions (C1) and (C2) hold. Let $\Gamma^* = [\gammab^*_j \mid j \in \NC \setminus \LC]$
              for the optimal solution $\gammab^*_j$ to $\AugProb_j(\LC, X^*)$,
              and let $\Pi$ be a permutation matrix of size $n$ such that $A \Pi = [A(\LC), A(\NC \setminus \LC)]$.
              Then,
              \begin{itemize}
                  \item $\opt(\HT(\LC, \LC))  = \opt(\HT(\NC, \NC))$, and
                  \item the matrix
                        \begin{align*}
                            \Pi
                            \begin{bmatrix}
                                X^* & \Gamma^* \\
                                O   & O
                            \end{bmatrix}
                            \Pi^\trans
                        \end{align*}
                        is the optimal solution to $\HT(\NC, \NC)$.
              \end{itemize}
    \end{enumerate}
\end{thm}
The proof is provided in Appendix~\ref{Appx: Proof of main theorem}.
Note that $((Y^*)^\trans \a_j)^+$ in condition (C2) is equivalent to $\b_j^+$ for $\b_j = (Y^*)^\trans \a_j$.
One might be concerned that condition (C2) rarely holds; i.e., it will not hold if $v^*$ is positive.
However, this concern is alleviated by Lemma \ref{Lem: sign of v}, which ensures that $v^*$ is always nonpositive.

\begin{lem} \label{Lem: sign of v}
    Let $(Y^*, Z^*, \s^*, \t^*, v^*)$ be the optimal solution to $\Dual(\LC,\LC)$.
    If $\LC$ is chosen to satisfy $\ell \ge r$, then $v^* \le 0$.
\end{lem}
The proof is provided in Appendix~\ref{Appx: Proof of sign of v}.
Algorithm~\ref{Alg: solving Hottopixx model} describes the RCE algorithm for computing the optimal solution to $\HT$.

\begin{algorithm}[h]
    \caption{RCE : Row and column expansion algorithm for computing the optimal solution to $\HT$} \label{Alg: solving Hottopixx model}
    \smallskip
    Input: $A \in \Real^{d \times n}$ and a positive integer $r$. \\
    Output: $X \in \Real^{n \times n}$.
    \begin{enumerate}[1:]
        \item Choose $\LC \subset \NC$ satisfying $r \le \ell$.
        \item Repeat the following procedure until the stopping condition at step 2-2 or 2-3 is met.
              \begin{enumerate}[{2-}1:]
                  \item Compute the optimal solution $(X^*, F^*, G^*, u^*)$ to $\Primal(\LC, \LC)$,
                        and the optimal solution $(Y^*, Z^*, \s^*, \t^*, v^*)$ to $\Dual(\LC, \LC)$.
                  \item If $\LC = \NC$, return $X = X^*$ and terminate; otherwise, go to step 2-3.
                  \item For each $j \in \NC \setminus \LC$,
                        compute the optimal solution $\gammab_j$ and the optimal value of $\AugProb_j(\LC, X^*)$.
                        If
                        \begin{align*}
                            \opt(\AugProb_j(\LC, X^*)) \le \opt(\Primal(\LC, \LC))
                        \end{align*}
                        for every $j \in \NC \setminus \LC $, go to step 3;
                        otherwise, update $\LC$ by
                        \begin{align*}
                            \LC \cup \{j \in \NC \setminus \LC \ | \ \opt(\AugProb_j(\LC, X^*)) > \opt(\Primal(\LC, \LC)) \},
                        \end{align*}
                        and go back to step 2-1.
              \end{enumerate}
        \item
              If
              \begin{align*}
                  v^* + \one^\trans ((Y^*)^\trans \a_j)^+ \le 0
              \end{align*}
              for every $j \in \NC \setminus \LC$, then construct the matrix,
              \begin{align*}
                  X = \Pi
                  \begin{bmatrix}
                      X^* & \Gamma^* \\
                      O   & O
                  \end{bmatrix}
                  \Pi^\trans  \in \Real^{n \times n}
              \end{align*}
              for $\Gamma^* = [\gammab_j^*  \mid  j \in \NC \setminus \LC]$ and a permutation matrix $\Pi$ such that $A \Pi = [A(\LC), A(\NC \setminus \LC)]$.
              Return $X$ and terminate.
              Otherwise, update $\LC$ by
              \begin{align*}
                  \LC \cup \{j \in \NC \setminus \LC \ | \ v^* + \one^\trans  ((Y^*)^\trans \a_j)^+ > 0 \},
              \end{align*}
              and go back to step 2.
    \end{enumerate}
\end{algorithm}

\subsection{Discussion on the Computational Cost of RCE} \label{Subsec: discussion on the computational cost of RCE}
RCE updates the index set $\LC$ iteratively,
and the number of elements in $\LC$, denoted by $\ell$, increases over the iterations.
In what follows, we examine how the computational cost of RCE depends on $\ell$.

Solving $\Primal(\LC, \LC)$ and $\Dual(\LC, \LC)$ in step 2-1 is the most computationally expensive part of RCE.
Assume that $\ell \ge d$.
When $\Primal(\LC, \LC)$ is converted into the standard form LP,
it involves $O(\ell^2)$ variables and $O(\ell^2)$ constraints.
As reviewed in Section \ref{Subsec: duality and solution methods for LPs},
when employing the primal-dual interior-point method to compute the optimal solutions of
$\Primal(\LC, \LC)$ and $\Dual(\LC, \LC)$,
the iteration count is $O( \ell \log(1/\epsilon))$ for a given tolerance $\epsilon > 0$,
and each iteration requires solving a system of linear equations with
$O(\ell^2)$ variables and $O(\ell^2)$ equations.
In step 2-3,
RCE must solve $\AugProb_j(\LC, X^*)$ for each $j \in \NC \setminus \LC$.
When $\AugProb_j(\LC, X^*)$ is converted into the standard form LP,
it involves $O(\ell)$ variables and $O(\ell)$ constraints,
which is considerably smaller in size than $\Primal(\LC, \LC)$.

From the above discussion, we see that
if $\ell$ remains sufficiently smaller than $n$ during the iteration process,
and the number of times $\Primal(\LC, \LC)$ and $\Dual(\LC, \LC)$ must be solved is small,
then RCE can solve $\HT$ within a reasonable amount of time, even when $n$ is large.
We provide numerical evidence supporting this claim in Section~\ref{Subsec: computational efficiency}.

\section{Detailed Description of EEHT} \label{Sec: detailed description of EEHT}
This section provides a detailed explanation of EEHT, in which RCE is a key component, as well as a summary of the overall algorithm.

\paragraph{Preprocessing}
Before running RCE,
we apply a dimensionality reduction technique to the input matrix.
Given $A \in \Real^{d \times n}$ and a positive integer $r$ as input,
we compute the top-$r$ truncated SVD,
\begin{align*}
    A_r = U_r \Sigma_r V_r^\trans,
\end{align*}
of $A$.
Here, $\Sigma_r \in \Real^{r \times r}$ is diagonal so that
the top-$r$ singular values $\sigma_1, \ldots, \sigma_r$ of $A$ are in the diagonal positions and
the columns of $U_r \in \Real^{d \times r}$ (resp. $V_r \in \Real^{n \times r}$) are
the left-singular (resp. right-singular) vectors of $A$
corresponding to $\sigma_1, \ldots, \sigma_r$.
Then, we construct a size-reduced matrix $A' = \Sigma_r V_r^\trans \in \Real^{r \times n}$ of $A$
and run RCE on input $(A', r)$.
It should be noted that if $A$ is an HSI matrix of size $d \times n$ with $r$ endmembers, as shown in (\ref{Eq: LMM under PPA}),
then its size-reduced matrix $A'$ is also an HSI matrix, written as $A' = W'[I, \bar{H}]\Pi + V'$
for $W' \in \Real^{r \times r}$ and $V' \in \Real^{r \times n}$.

\paragraph{Step 2-2 of Algorithm \ref{Alg: HT}} \label{Para: choice of element}
In theory, we are allowed to choose any one of the elements in a cluster $\SC_\ell$ at step 2-2 of Algorithm \ref{Alg: HT}.
Indeed, Theorem 3.2 of \cite{Miz22} ensures that Algorithm \ref{Alg: HT} is robust to noise,
and this result holds no matter how we choose an element from $\SC_\ell$ at step 2-2.
Yet, in practice, the method of choice affects its robustness to noise.
For that reason, we employ two methods of choice in step 2-2 of Algorithm \ref{Alg: HT}.
The first one, called the {\it max-point choice}, chooses an element $u \in \SC_\ell$ with the highest score
according to the point list $\p_\ell$:
\begin{align} \label{Eq: max-point choice}
    u = \arg \max_{u \in \SC_\ell} \p_\ell(u).
\end{align}
The max-point choice is not new; it was used in \cite{Gil13, Gil14b, Miz22} in their postprocessing for Hottopixx.
The second one, called the {\it cluster centroid choice}, defines the centroid $\c_\ell$ of a cluster $\SC_\ell$ by
\begin{align*}
    \c_\ell = \frac{1}{|\SC_\ell|}\sum_{u \in \SC_\ell} \a_u
\end{align*}
and then chooses an element $u \in \SC_\ell$ such that $\a_u$ is closest to $\c_\ell$ in terms of MRSA:
\begin{align} \label{Eq: cluster centroid choice}
    u = \arg \min_{u \in \SC_\ell} \MRSA(\c_\ell, \a_u).
\end{align}
The cluster centroid choice is our proposal.

\paragraph{Step 1 of RCE}
The initial choice of $\LC$ at step 1 of RCE can affect the growth of $\LC$ during the iterations.
If the initial $\LC$ includes many column indices of $A$ corresponding to pure pixels,
the subsequent growth of $\LC$ can be suppressed.
We employ SPA \cite{Ara01, Gil14a} for this purpose,
as SPA is fast and its output is reasonably accurate.
Algorithm \ref{Alg: construction of initial index set} describes our procedure for constructing an initial set $\LC$.
Step 1 uses SPA to find columns $\a_{\ell_1}, \ldots, \a_{\ell_r}$ of $A$ that are expected to be close to pure pixels.
Step 2 finds the top-$\zeta$ nearest neighbors of each $\a_{\ell_1}, \ldots, \a_{\ell_r}$
and constructs a set $\LC_{\mathrm{NN}}$ of their indices.
Step 3 chooses $\eta$ extra elements from the complement of $\LC_{\mathrm{NN}}$
and constructs a set $\LC_{\mathrm{EX}}$ of the extras.
Step 4 takes the union of $\LC_{\mathrm{NN}}$ and $\LC_{\mathrm{EX}}$ and returns it as $\LC$.

Algorithm \ref{Alg: implementation of Hottopixx} is the overall procedure of EEHT.

\begin{algorithm}[h]
    \caption{Construct an initial index set $\LC$} \label{Alg: construction of initial index set}
    \smallskip
    Input: $A \in \Real^{d \times n}$ and  positive integers $r, \zeta$ and $\eta$. \\
    Output: $\LC \subset \NC$.
    \begin{enumerate}[1:]
        \item Perform SPA on input $(A, r)$. Let $\a_{\ell_1}, \ldots, \a_{\ell_r}$  be the columns of $A$ output by it.
        \item Set $\LC_{\mathrm{NN}} = \emptyset$. Perform the following procedure for $i = 1, \ldots, r$.
              \begin{enumerate}[{2-}1:]
                  \item Sort columns $\a_1, \ldots, \a_n$ of $A$ by their distance to $\a_{\ell_i}$ in ascending order
                        so that
                        \begin{align*}
                            \|\a_{\ell_i} - \a_{u_1} \|_2 \le \cdots \le \|\a_{\ell_i} - \a_{u_n} \|_2
                        \end{align*}
                        where $\{u_1, \ldots, u_n\} = \NC$.
                  \item Update $\LC_{\mathrm{NN}}$ by $\LC_{\mathrm{NN}} \cup \{u_1, \ldots, u_{\zeta} \}$.
              \end{enumerate}
        \item Choose $\eta$ arbitrary elements from $\NC \setminus \LC_{\mathrm{NN}}$,
              and construct a set $\LC_{\mathrm{EX}}$ of them.
        \item Set $\LC = \LC_{\mathrm{NN}} \cup \LC_{\mathrm{EX}}$ and return $\LC$.
    \end{enumerate}
\end{algorithm}

\begin{algorithm}[h]
    \caption{EEHT: Efficient and effective implementation of Hottopixx} \label{Alg: implementation of Hottopixx}
    \smallskip
    Input: $A \in \Real^{d \times n}$ and positive integers $r, \zeta$, and $\eta$. \\
    Output: $r$ columns of $A$.
    \begin{enumerate}[1:]
        \item Compute the top-$r$ truncated SVD $A_r = U_r \Sigma_r V_r^\trans$
              and construct a size-reduced matrix $A' = \Sigma_r V_r^\trans$.
        \item Run RCE on $(A', r)$ and obtain the output $X$,
              where step 1 runs Algorithm \ref{Alg: construction of initial index set} on $(A', r, \zeta, \eta)$.
        \item Construct an index set $\IC$ by performing one of the following methods A-C and return $\a_i$ with $i \in \IC$.
              \begin{enumerate}[A.]
                  \item Find a set $\IC$ of $r$ indices corresponding to the $r$ largest elements of $\diag(X)$,
                        and return $\a_i$ with $i \in \IC$.
                  \item	Run step 2 of Algorithm \ref{Alg: HT}
                        where step 2-2 adopts the max-point choice of (\ref{Eq: max-point choice}).
                  \item Run step 2 of Algorithm \ref{Alg: HT}
                        where step 2-2 adopts the cluster centroid choice of (\ref{Eq: cluster centroid choice}).
              \end{enumerate}
    \end{enumerate}
\end{algorithm}

\section{Experiments} \label{Sec: experiments}
We conducted experiments to assess the efficiency and effectiveness of EEHT on Problem \ref{Prob: endmember extraction}.
For the experiments, we coded EEHT in MATLAB.
Here, step 2 of EEHT runs RCE whose step 2-1 needs to solve LP problems $\Primal(\LC, \LC)$ and $\Dual(\LC, \LC)$.
We employed CPLEX for solving these LP problems.
The MATLAB function {\tt cplexlp} is available in the CPLEX package and it enables us to run CPLEX in the MATLAB environment.
Applying the {\tt cplexlp} function to an LP returns both the optimal solutions for the LP and its dual.
Our code used {\tt cplexlp} for $\Dual(\LC, \LC)$
in order to obtain optimal solutions to both $\Primal(\LC, \LC)$ and $\Dual(\LC, \LC)$.
In what follows, EEHT with method A (resp. B, C) chosen at step 3 is referred to as EEHT-A (resp. -B, -C).

We used MRSA to assess the endmember extraction performance.
Given the columns $\a_{i_1}, \ldots, \a_{i_r}$ of $A$ returned by the algorithm,
each estimated endmember signature $\hat{\w}_j$ is defined as $\hat{\w}_j = \a_{i_j}$.
Let $\hat{\w}_1, \ldots, \hat{\w}_r$ be the estimated endmember signatures, and
$\w_1, \ldots, \w_r$ be the reference signatures.
We computed a permutation $\sigma$ of size $r$ such that
\begin{align*}
    \min_{\sigma} \sum_{j=1}^{r} \MRSA(\hat{\w}_j, \w_{\sigma(j)}).
\end{align*}
Note that the minimum is taken over all permutations $\sigma$ of size $r$.
We then evaluated the MRSA values between the estimated endmember signatures and the reference ones,
\begin{align*}
    \MRSA(\hat{\w}_j, \w_{\sigma(j)}) \ \text{for} \ j = 1, \ldots, r
\end{align*}
and their average,
\begin{align*}
    \frac{1}{r} \sum_{j=1}^{r} \MRSA(\hat{\w}_j, \w_{\sigma(j)}).
\end{align*}
We refer to this average as the \textit{MRSA score}, which represents endmember extraction performance.
The experiments were done in MATLAB on dual Intel Xeon Gold 6336Y processors with 256 GB of memory.

\subsection{Computational Efficiency} \label{Subsec: computational efficiency}
RCE is a key component of EEHT in solving Hottopixx models efficiently.
We thus examined the computational time of RCE on synthetic datasets.
The datasets contained HSI matrices $A = W [I, \bar{H}] + V$ with $r$ endmembers
that were generated by the following procedure:
\begin{itemize}

    \item \textit{As for $W$:}
          Draw the entries of $W$ from a uniform distribution on the interval $[0,1]$,
          and replace them with their absolute values.
          Finally, normalize the columns of $W$ to have a unit $L_1$ norm.
    \item \textit{As for $\bar{H}$:}
          Draw the columns of $\bar{H}$ from a Dirichlet distribution
          with $r$ parameters that are uniformly distributed in the interval $[0,1]$.
    \item \textit{As for $V$:}
          Draw the entries of $V$ from a standard normal distribution,
          and normalize $V$ such that $\nu = \|V\|_1$ for a noise intensity level $\nu$ specified in advance.
\end{itemize}
We chose ten equally spaced points between $0$ and $1$ and set them as noise intensity levels $\nu$.
Let $\nu_1, \ldots, \nu_{10}$ denote these levels in increasing order, so that
$0 = \nu_1 \le \cdots \le \nu_{10} = 1$.
Each dataset contained ten HSI matrices $A = W [I, \bar{H}] + V$ of size $50 \times n$ with $r=10$
satisfying $\nu_i = \| V \|_1$ for each $i =1 ,\ldots, 10$.
We set $n$ from $500$ to $2500$ in $500$ increments and, thus, constructed five datasets in total.

We examined four methods, RCE-SR, RCE-DIR, CPLEX-SR, and CPLEX-DIR.
Here, RCE-DIR and CPLEX-DIR solved the Hottopixx models $\HT$ for the original matrices $A$,
while RCE-SR and CPLEX-SR solved theirs for size-reduced matrices $A'$ obtained
from the top-$r$ truncated SVD of $A$.
The details are as follows.
\begin{itemize}
    \item \textbf{RCE-SR} constructed a size-reduced matrix $A'$ of $A$
          and then applied RCE to $A'$,
          where step 1 runs Algorithm \ref{Alg: construction of initial index set} with $(\zeta, \eta) = (10, 100)$.
    \item \textbf{RCE-DIR} directly applied RCE to $A$,
          where step 1 runs Algorithm \ref{Alg: construction of initial index set} with $(\zeta, \eta) = (10, 100)$.
    \item \textbf{CPLEX-SR} constructed a size-reduced matrix $A'$ of $A$
          and used \texttt{cplexlp} to solve the LP problem $\Primal(\NC, \NC)$ for $A'$.
    \item \textbf{CPLEX-DIR} directly used \texttt{cplexlp} to solve the LP problem $\Primal(\NC, \NC)$ for $A$.
\end{itemize}

\begin{figure}[h]
    \centering
    \includegraphics[width=0.8\linewidth]{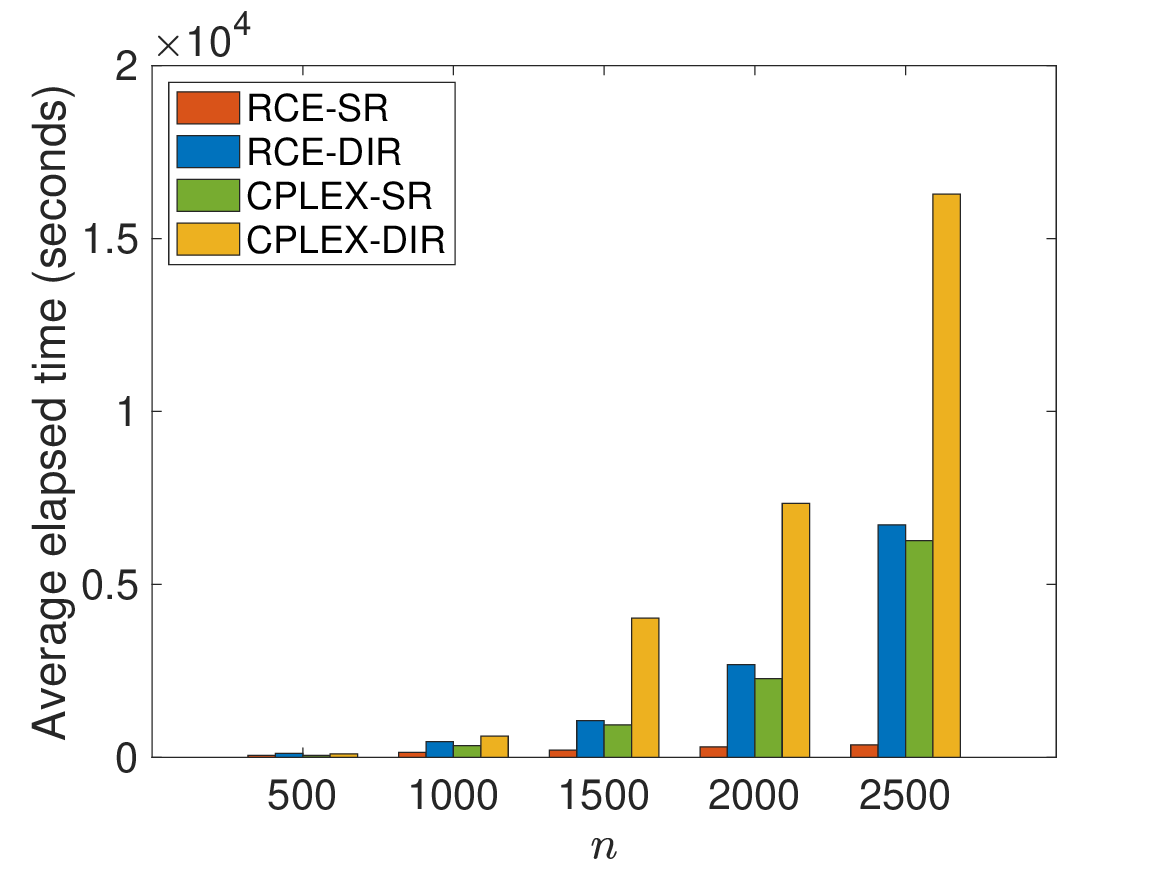}
    \caption{Average elapsed time (in seconds) of four methods for five datasets.}
    \label{Fig: exp 1 - average elapsed time}
\end{figure}

We compared RCE-SR and -DIR with CPLEX-SR and -DIR,
since the Hottopixx methods in \cite{Gil13, Gil14b, Miz22} use CPLEX and solve the Hottopixx models directly.
Figure \ref{Fig: exp 1 - average elapsed time} summarizes the average elapsed time for each dataset.
We can see from the figure that RCE is faster than CPLEX for $n \ge 1000$.
In particular, RCE-SR is significantly faster than the other methods
and its average elapsed time increases more slowly.
RCE-DIR is faster than CPLEX-DIR, but its average elapsed time increases more rapidly than that of RCE-SR.
The results suggest that RCE-SR should be used for implementing Hottopixx
when dealing with large matrices.

To better understand why RCE-SR was faster than the other methods,
we provide Tables \ref{Tab: exp 1 - detailed results of RCE-DIR and CPLEX-DIR} and \ref{Tab: exp 1 - detailed results of RCE-SR and CPLEX-SR},
which summarize the results for ten HSI matrices with varied noise intensity levels $\nu_1, \ldots, \nu_{10}$ in the dataset with $n=2500$.
For RCE-SR,
Table \ref{Tab: exp 1 - detailed results of RCE-SR and CPLEX-SR} shows that
$\ell$ did not increase significantly during the iterations,
which in turn led to a substantial reduction in computation time compared with CPLEX-SR.
For RCE-DIR,
Table \ref{Tab: exp 1 - detailed results of RCE-DIR and CPLEX-DIR} indicates that
the maximum values of $\ell$ exceeded $n/2$ for $\nu_5, \ldots, \nu_{10}$
and in particular, that it solved $\Primal(\LC, \LC)$ and $\Dual(\LC, \LC)$
four times instead of three for $\nu_5$ and $\nu_9$.
Consequently, the elapsed times of RCE-DIR were nearly the same as those of CPLEX-DIR for $\nu_5$ and $\nu_9$.

We also examined the effect of the parameters $\zeta$ and $\eta$ on the computational time of RCE-SR.
In the experiments,
$\zeta$ was varied from $5$ to $20$ in increments of $5$, and $\eta$ was varied from $50$ to $200$ in increments of $50$,
resulting in $16$ parameter combinations.
For each combination of $(\zeta, \eta)$, RCE-SR was applied to six datasets,
comprising the five datasets used in the previous experiments and an additional dataset with $n=5000$.
The results indicated that the elapsed time of RCE-SR did not change substantially
when $(\zeta, \eta)$ were varied across the $16$ combinations.
Detailed experimental results are provided in Appendix \ref{Appx: Effect of parameters zeta and eta on elapsed time of RCE-SR}.

\begin{table*}[t!]
    \centering
    \caption{Comparison of RCE-DIR and CPLEX-DIR on ten HSI matrices with noise intensity levels
        $\nu_1, \ldots, \nu_{10}$ in the dataset with $n=2500$.
        The table reports the elapsed time (top block), the maximum value of $\ell$ observed during the iterations (middle block),
        and the number of times $\Primal(\LC,\LC)$ and $\Dual(\LC,\LC)$ were solved (bottom block).
        Entries correspond to $\nu_1, \ldots, \nu_{10}$, ordered from left to right.}
    \label{Tab: exp 1 - detailed results of RCE-DIR and CPLEX-DIR}
    \begin{tabular}{ r r r r r r r r r r }
        \hline
        \multicolumn{10}{c}{Elapsed time in seconds ($\times 10^{-2}$): RCE-DIR (top) and CPLEX-DIR (bottom)} \\
        \hline
        1.2  & 6.5   & 8.9   & 50.1  & 110.9 & 173.0 & 97.4  & 85.4  & 99.9 & 38.6                            \\
        69.0 & 444.8 & 174.4 & 119.5 & 121.5 & 214.2 & 169.7 & 158.7 & 92.1 & 64.3                            \\
        \hline
        \addlinespace
        \hline
        \multicolumn{10}{c}{Maximum value of $\ell$ during the iterations of RCE-DIR}                         \\
        \hline
        200  & 424   & 709   & 995   & 1256  & 1340  & 1525  & 1614  & 1630 & 1487                            \\
        \hline
        \addlinespace
        \hline
        \multicolumn{10}{c}{Number of times RCE-DIR solved $\Primal(\LC,\LC)$ and $\Dual(\LC,\LC)$}           \\
        \hline
        1    & 4     & 3     & 3     & 4     & 3     & 3     & 3     & 4    & 3                               \\
        \hline
    \end{tabular}
\end{table*}
\begin{table*}[t!]
    \centering
    \caption{Comparison of RCE-SR and CPLEX-SR on ten HSI matrices with noise intensity levels
        $\nu_1, \ldots, \nu_{10}$ in the dataset with $n=2500$.
        The table reports the elapsed time (top block), the maximum value of $\ell$ observed during the iterations (middle block),
        and the number of times $\Primal(\LC,\LC)$ and $\Dual(\LC,\LC)$ were solved (bottom block).
        Entries correspond to $\nu_1, \ldots, \nu_{10}$, ordered from left to right.}
    \label{Tab: exp 1 - detailed results of RCE-SR and CPLEX-SR}
    \begin{tabular}{ r r r r r r r r r r }
        \hline
        \multicolumn{10}{c}{Elapsed time in seconds ($\times 10^{-2}$): RCE-SR (top) and CPLEX-SR (bottom)} \\
        \hline
        1.1  & 3.4   & 3.4   & 3.4  & 5.8  & 4.2  & 3.0  & 3.6  & 5.1  & 3.2                                \\
        31.9 & 109.9 & 175.8 & 52.5 & 50.2 & 48.2 & 38.0 & 39.4 & 36.5 & 44.0                               \\
        \hline
        \addlinespace
        \hline
        \multicolumn{10}{c}{Maximum value of $\ell$ during the iterations of RCE-SR}                        \\
        \hline
        200  & 239   & 276   & 308  & 516  & 517  & 615  & 735  & 673  & 663                                \\
        \hline
        \addlinespace
        \hline
        \multicolumn{10}{c}{Number of times RCE-SR solved $\Primal(\LC,\LC)$ and $\Dual(\LC,\LC)$}          \\
        \hline
        1    & 3     & 3     & 3    & 4    & 3    & 2    & 2    & 3    & 2                                  \\
        \hline
    \end{tabular}
\end{table*}

\subsection{Endmember Extraction Performance} \label{Subsec: endmember extraction performance}
\begin{figure}[h]
    \centering
    \begin{minipage}[b]{0.3\linewidth}
        \includegraphics[width=1.0\linewidth]{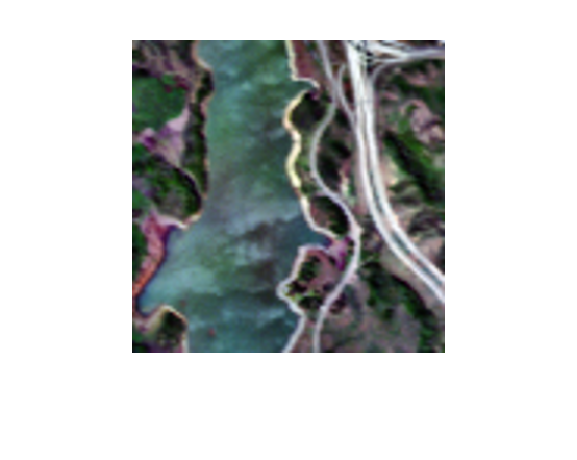}
    \end{minipage}
    \begin{minipage}[b]{0.3\linewidth}
        \includegraphics[width=1.0\linewidth]{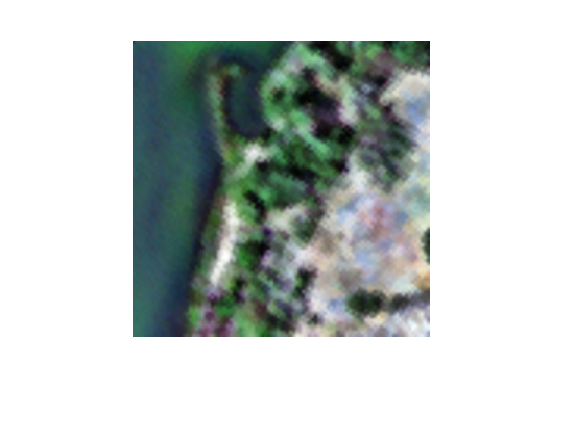}
    \end{minipage}
    \caption{RGB images of Jasper Ridge (left) and Samson (right).}
    \label{Fig: RGB images of jasper and samson}
\end{figure}

We examined the endmember extraction performance of EEHT using synthetic HSI datasets.
Experiments were first conducted on LMMs and subsequently on bilinear mixing models (BMMs)
to assess the capability of EEHT in the presence of multiple photon interactions.
BMMs have attracted attention as models for the nonlinear unmixing of hyperspectral images;
see, for example,~\cite{Yan18a, Yan18b, Gu22}.
The datasets were derived from the Jasper Ridge and Samson HSI datasets,
and subimages were extracted following the procedure described in Section~V of \cite{Zhu14}:
\begin{itemize}
    \item Jasper Ridge consists of $100 \times 100$ pixels with $198$ bands and contains $4$ endmembers: Tree, Soil, Water, and Road.
    \item Samson consists of $95 \times 95$ pixels with $156$ bands and contains $3$ endmembers: Soil, Tree, and Water.
\end{itemize}
Figure \ref{Fig: RGB images of jasper and samson} displays RGB images of these datasets.
The endmember signatures in these regions, previously identified in \cite{Zhu14},
were used as reference signatures for generating the synthetic datasets.

First, we evaluated the performance of EEHT on synthetic datasets generated from LMMs.
Let $A^\real$ be the HSI matrix for a dataset,
and $\w^\reference_1, \ldots, \w^\reference_r$ be the reference signature vectors.
The following procedure was applied:
\begin{enumerate}[1:]
    \item Normalize all columns of $A^\real$ to have a unit $L_1$ norm.
    \item For $i=1, \ldots, r$, find
          \begin{align*}
              j_i = \arg \min_{j =1, \ldots, n} \MRSA(\a^\real_j, \w^\reference_i),
          \end{align*}
          and then construct $\JC = \{j_1, \ldots, j_r\}$.
    \item Set $W = A^\real(:, \JC)$.
    \item Compute the optimal solution $X$ to problem (\ref{Prob: abundance computation}) by letting $(A, \IC) = (A^\real,\JC)$ in the problem.
          Set $X(:, \JC) = I$ and then $H = X$.
    \item Set $V = A^\real - WH$.
\end{enumerate}

We constructed two datasets using the $W, H$ and $V$ obtained above:
dataset 1 from Jasper Ridge and dataset 2 from Samson.
Each dataset contained $20$ HSI matrices that were generated as follows:
choose $20$ equally spaced points between $0$ and $1$ as noise intensity levels $\nu$;
and construct an HSI matrix $A = WH + (\nu / \|V\|_1) \cdot V$ for each $\nu$.
The HSI matrices $A$ were of size $198 \times 10000$ with $4$ endmembers for dataset 1
and of size $156 \times 9025$ with $3$ endmembers for dataset 2.
Note that $A = A^\real$ holds if $\nu = \| V \|_1$,
where $\| V \|_1 \approx 0.61$ for dataset 1 and $\| V \|_1 \approx 0.15$ for dataset 2.

\begin{figure*}[t!]
    \centering
    \subfloat[Results of EEHT-A, -B, -C on dataset 1.]{
        \includegraphics[width=0.3\linewidth]{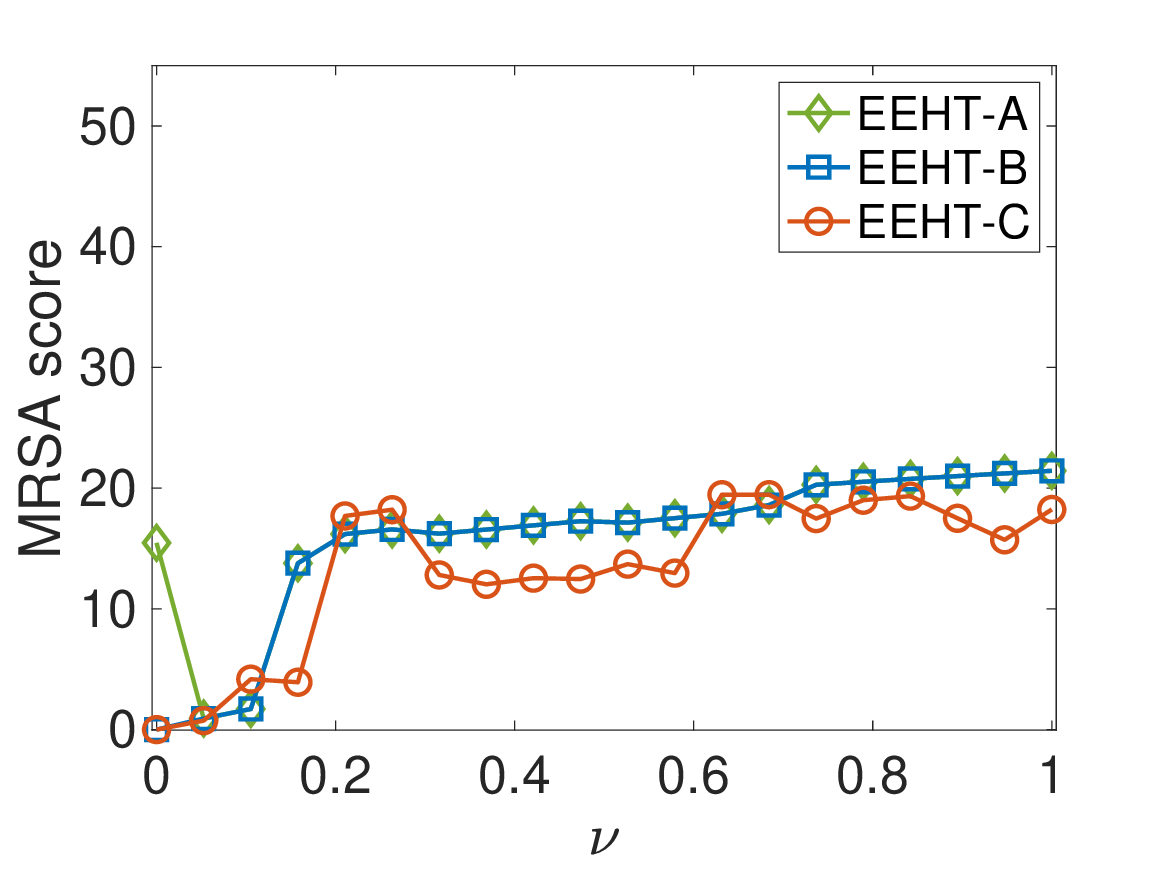}
    }
    \hfil
    \subfloat[Results of EEHT-C and the convex method MERIT on dataset 1.]{
        \includegraphics[width=0.3\linewidth]{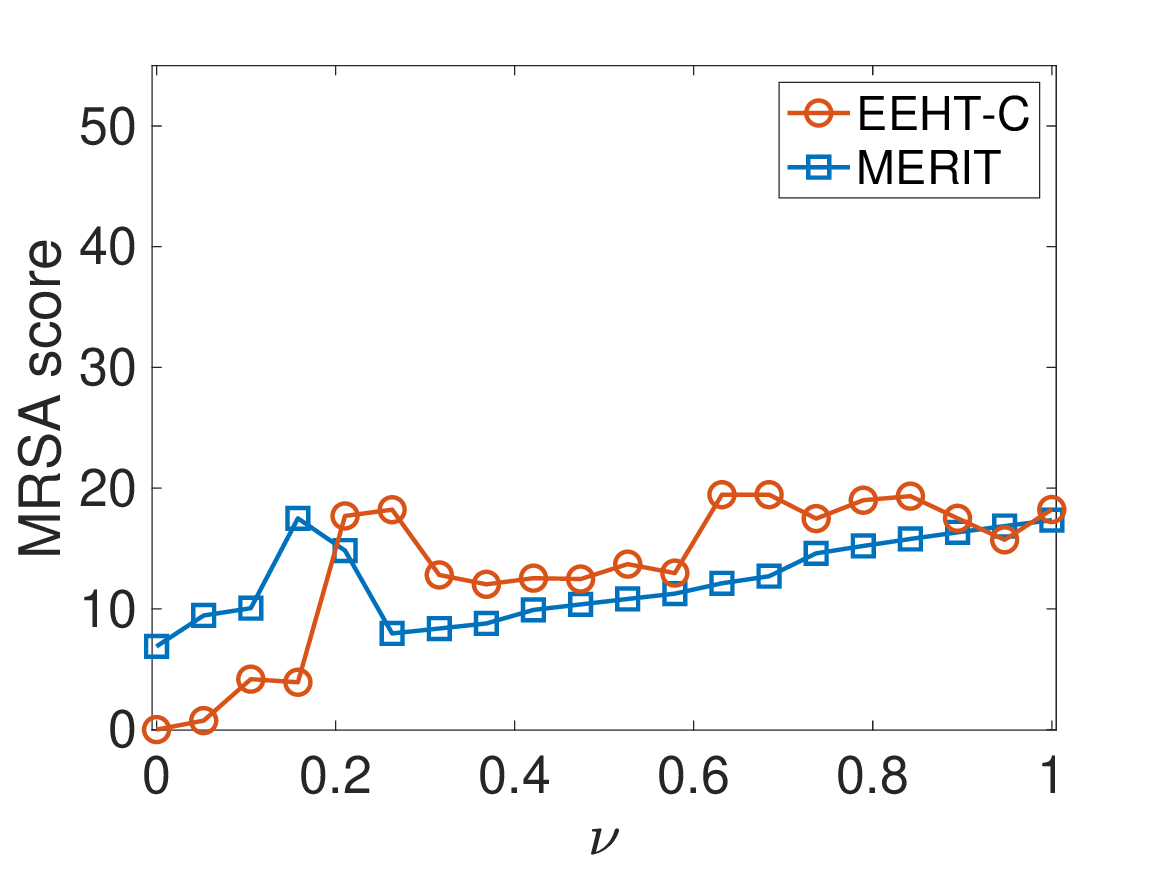}
    }
    \hfil
    \subfloat[Results of EEHT-C and  the greedy methods SPA, PSPA, ER, VCA, and SNPA on dataset 1.]{
        \includegraphics[width=0.3\linewidth]{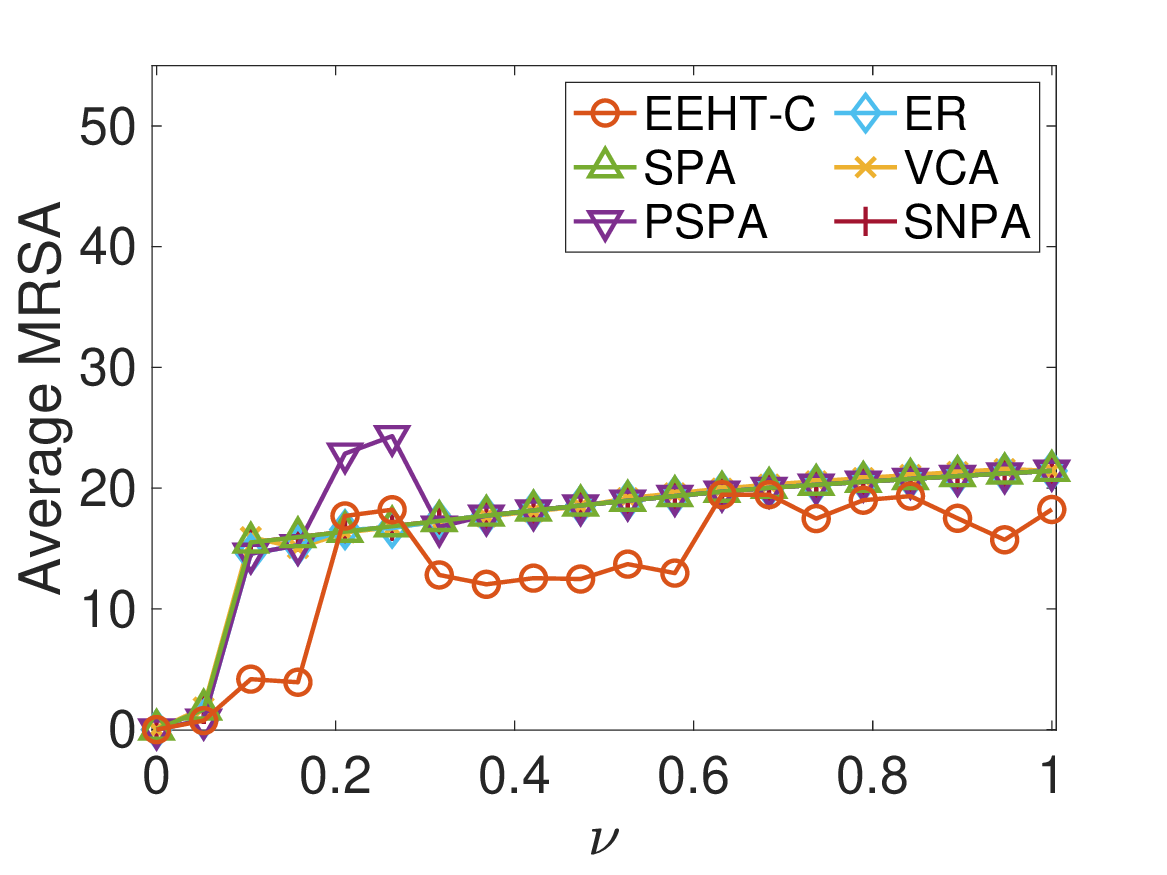}
    }
    \vfil
    \subfloat[Results of EEHT-A, -B, -C on dataset 2.]{
        \includegraphics[width=0.3\linewidth]{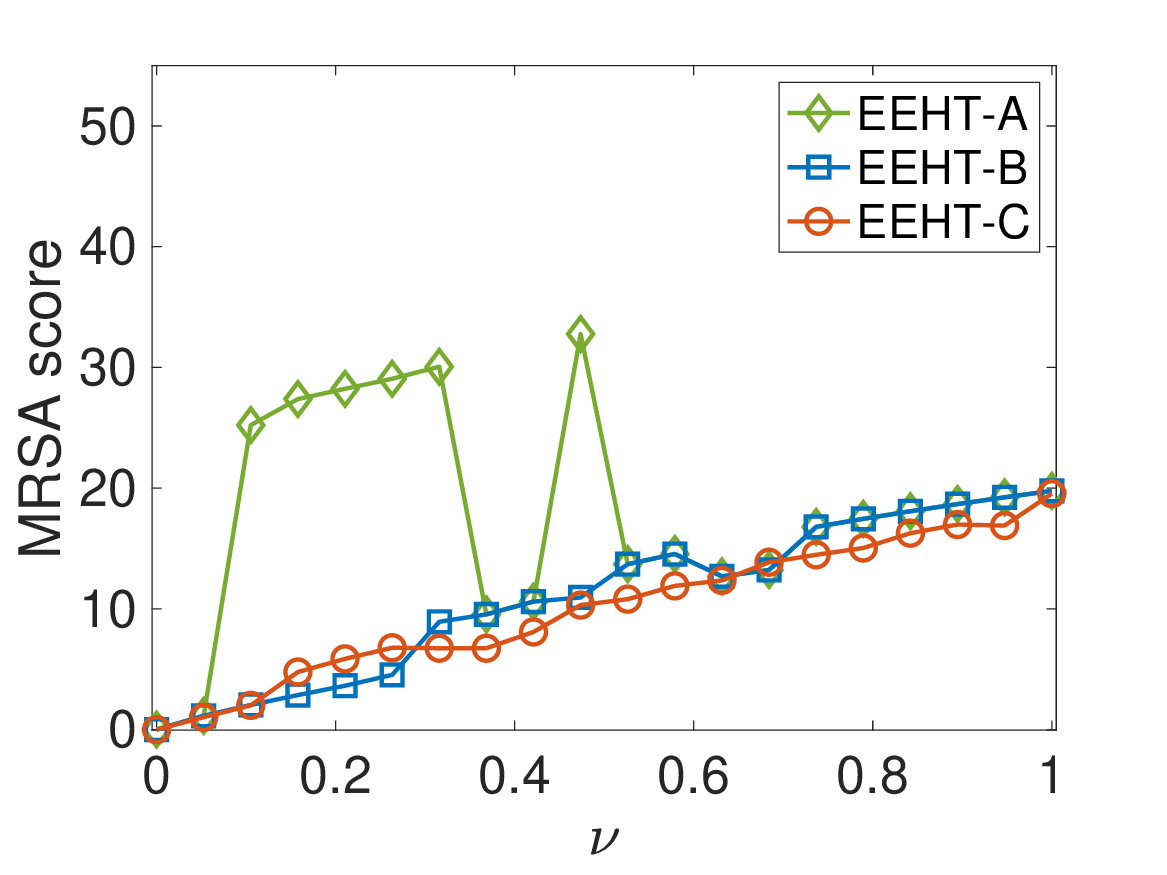}
    }
    \hfil
    \subfloat[Results of EEHT-C and the convex method MERIT on dataset 2.]{
        \includegraphics[width=0.3\linewidth]{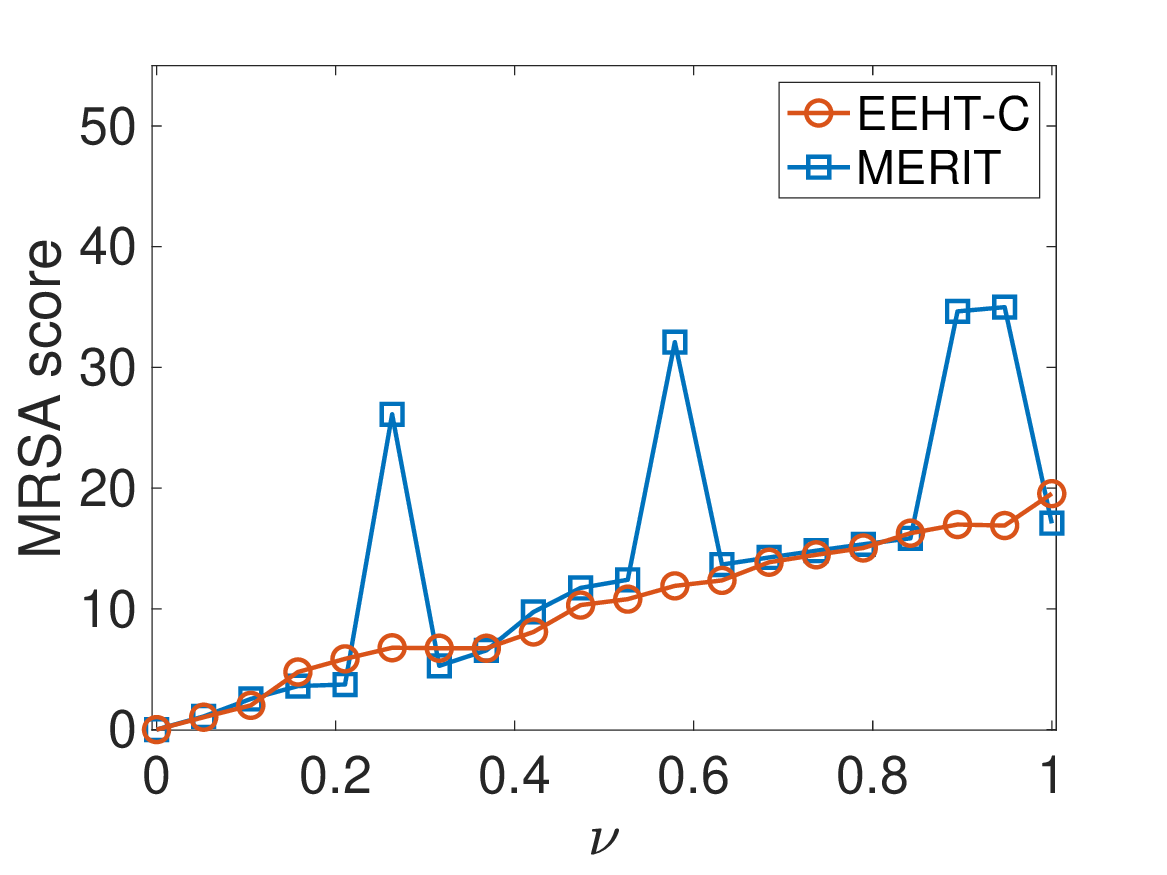}
    }
    \hfil
    \subfloat[Results of EEHT-C and the greedy methods SPA, PSPA, ER, VCA, and SNPA on dataset 2.]{
        \includegraphics[width=0.3\linewidth]{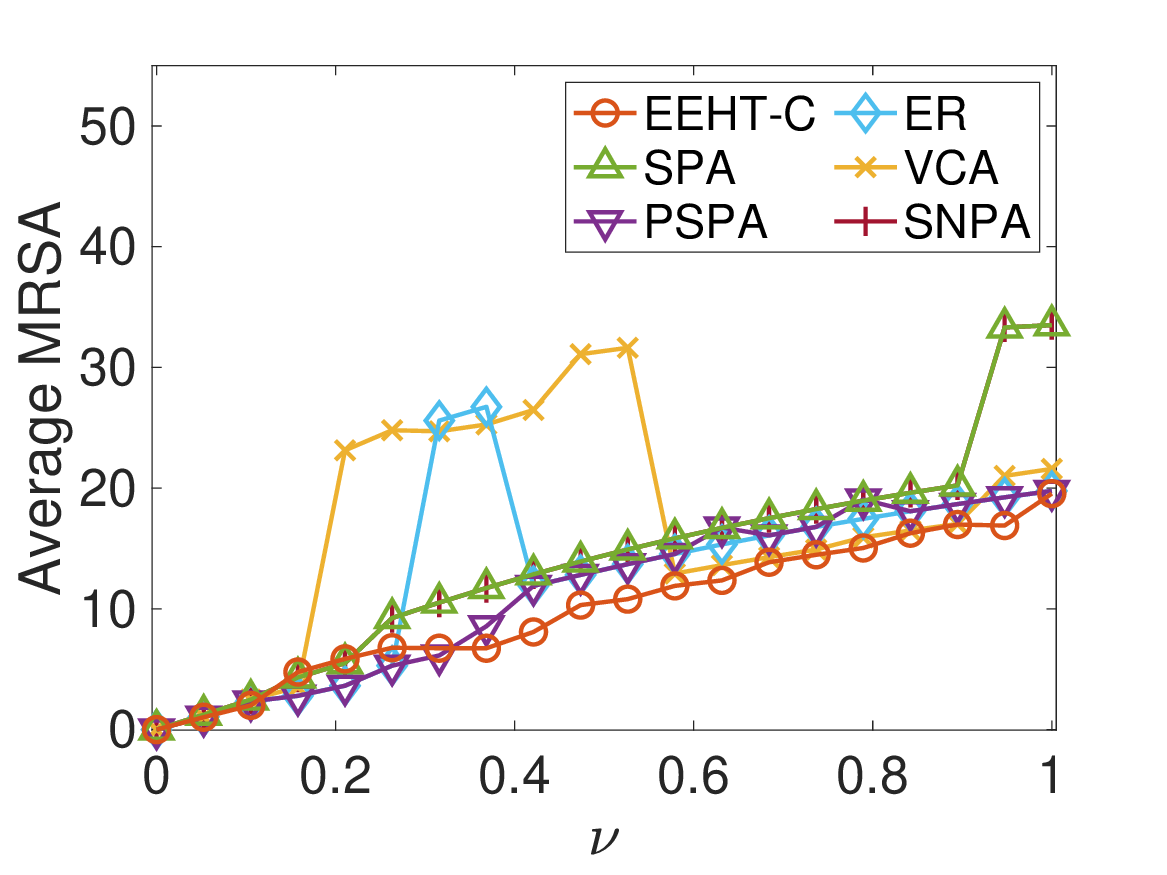}
    }
    \caption{MRSA scores ($\times 10^2$) of EEHT and existing methods for dataset 1 (top) and dataset 2 (bottom).}
    \label{Fig: exp 2}
\end{figure*}

We compared EEHT with six existing methods: MERIT, SPA, PSPA, ER, VCA, and SNPA.
While MERIT, like EEHT, is a convex optimization-based method, the others are greedy methods.
We developed MATLAB codes for SPA, PSPA and ER and
used publicly available codes for MERIT, VCA, and SNPA, which were developed by the authors of \cite{Ngu22, Nas05, Gil14c}.
We set the input parameters $\zeta$ and $\eta$ of EEHT as $(\zeta, \eta) = (10,100)$.

For a fair comparison with EEHT, we included the SVD-based dimensionality reduction,
described in Section \ref{Sec: detailed description of EEHT}, in MERIT: i.e.,
first compute the top-$r$ truncated SVD $A_r = U_r \Sigma_r V_r^\top$ of the input matrix $A$;
then run MERIT on $A' = \Sigma_r V_r^\top$.

We ran EEHT and the six existing methods on datasets 1 and 2,
and evaluated their performance using MRSA scores.
Regarding MERIT, it is necessary to fix the parameter $\lambda$ of problem (\ref{Prob: self-dictionary sparse regression}) and
the parameter $\mu$ of the function $\phi_\mu$ shown in (\ref{Eq: phi}).
In particular, the choice of $\lambda$ affects its endmember extraction performance.
We set $\mu$ to the default value of $10^{-5}$, as suggested by the authors of \cite{Ngu22}.
We then set $\lambda$ to $10^{-6}, 10^{-5}, ..., 10^{3}$ and ran MERIT for each of these 10 values.
For each $\lambda$ value, we calculated the mean MRSA score across all intensity levels $\nu$.
For dataset 1 (resp., dataset 2), the mean score was minimized when $\lambda = 10^{2}$ (resp., $10^{-1}$).
Therefore, we report the results of MERIT with $\lambda = 10^{2}$ for dataset 1 and $\lambda = 10^{-1}$ for dataset 2 in Figure \ref{Fig: exp 2}.

The experiments were conducted by setting the parameters of the MERIT code  \cite{Ngu22} as follows.
The value of  $\lambda$  can be specified using the {\tt options.lambda} parameter, which we set as described above.
Additionally, we set {\tt options.backend} to {\tt mex}.
Default values were used for the remaining parameters.

Figure \ref{Fig: exp 2} plots the MRSA scores of the methods for datasets 1 and 2.
We can see from the figure that, in most cases, EEHT-C achieved better MRSA scores than EEHT-A, -B, and the five greedy methods.
This result suggests that using EEHT-C is preferable to EEHT-A and -B.
EEHT-C achieved worse MRSA scores than MERIT for dataset 1, while it performed better than MERIT for dataset 2.
For the $20$ HSI matrices included in each dataset, the computation times of EEHT-A, -B and -C were approximately 30 minutes on average.

\begin{remark}
    We ran FGNSR with an SVD-based dimensionality reduction	on datasets 1 and 2. It took about two weeks for each dataset.
    When running for multiple parameter values of $\lambda$, the computational time was excessively long.
    Thus, we did not conduct experiments on FGNSR.
\end{remark}

\begin{remark}
    Let us go back to the results of EEHT-A, -B, and -C on dataset 1 (top-left of Figure \ref{Fig: exp 2}).
    One may wonder why  EEHT-A took positive MRSA values at $\nu = 0$.
    Remark 4.1 of \cite{Miz22} predicts that
    this can happen if overlapping columns exist in the HSI matrices.
    Indeed, EEHT-A took a zero MRSA value for every endmember at $\nu = 0$
    after eliminating the duplicate columns.
\end{remark}

Next, we examined the performance of the methods tested in the LMM experiments,
when the effect of multiple photon interactions was taken into account in datasets 1 and 2.
This phenomenon is often observed when monitoring forested areas~\cite{Dob14, Yok14}.
In such areas, the interfaces between the canopy and the ground form a layered structure,
and light can be reflected by multiple materials before reaching a hyperspectral camera.

We assumed that the HSI data follow the generalized bilinear model (GBM) described in~\cite{Hal11}.
In the GBM, second-order photon scattering is accounted for by incorporating an additional term into the LMM.
That is, the observed spectral signature $\a_j$ at the $j$th pixel is expressed as
\begin{align*}
    \a_j = \sum_{i=1}^{r} h_{ij} \w_i
    + \sum_{p=1}^{r-1} \sum_{q=p+1}^{r} \xi_{pqj} h_{pj} h_{qj} \w_p \odot \w_q  + \v_j
\end{align*}
where $0 \le \xi_{pqj} \le 1$ and $\odot$ denotes the Hadamard product.
For the sake of simplicity,
let $\v_j'$ denote the term $\sum_{p=1}^{r-1} \sum_{q=p+1}^{r} \xi_{pqj} h_{pj} h_{qj} \w_p \odot \w_q$.
We can then write $A = [\a_1, \ldots, \a_n]$ as
\begin{align*}
    A = WH + V' + V
\end{align*}
where $V' = [\v'_1, \ldots, \v'_n]$. Note that $V'$ is determined by $W, H$, and $\xi_{pqj}$.

We synthetically generated two datasets based on GBMs: dataset 3 from Jasper Ridge and dataset 4 from Samson.
Each dataset contained HSI matrices defined by $A = WH + (\nu' / \| V' \|_1) \cdot V' + (\nu / \| V \|_1) \cdot V$
where $\nu$ is the noise intensity and $\nu'$ is the interaction intensity.
Dataset 3 used $W$, $H$, and $V$ obtained in the LMM experiments from Jasper Ridge,
while dataset 4 used those from Samson.
To construct $V'$, we used $W$ and $H$ and drew $\xi_{pqj}$ from a uniform distribution on the interval $[0,1]$.
For each dataset, we set $\nu' = 0.2$
and varied $\nu$ from $0.2$ to $0.8$ in increments of $0.2$.
For each $\nu$, we generated $20$ HSI matrices by changing $V'$,
resulting in $80$ HSI matrices in total per dataset.

For datasets 3 and 4,
we ran the methods with the same parameter settings as in the LMM experiments.
To avoid ambiguity, we specify that the MERIT parameter $\lambda$ was set to $10^{2}$ for dataset 3 and $10^{-1}$ for dataset 4.
Table~\ref{Tab: exp 2 - mean MRSA scores for datasets 3 and 4} summarizes the mean MRSA scores of EEHT and the six existing methods
for these datasets.
The results show that EEHT-C achieved better MRSA scores than the other methods for both datasets
when $\nu \ge 0.4$.

\begin{table}[h]
    \centering
    \caption{Mean MRSA scores ($\times 10^2$) for each noise intensity level $\nu$,
        averaged over 20 randomly generated HSI matrices at each level.
        Results for dataset 3 are presented in the first block, and those for dataset 4 in the second.
        The best scores are highlighted in bold.}
    \label{Tab: exp 2 - mean MRSA scores for datasets 3 and 4}
    \begin{tabular}{l r r r r }
        \hline
        \multicolumn{5}{c}{Dataset 3}                                         \\
        \hline
               & $\nu = 0.2$  & $\nu = 0.4$   & $\nu =0.6$    & $\nu = 0.8$   \\
        \hline
        EEHT-A & 16.4         & 19.5          & 20.3          & 20.0          \\
        EEHT-B & 14.1         & 16.7          & 18.2          & 20.1          \\
        EEHT-C & 14.1         & \textbf{16.1} & \textbf{14.5} & \textbf{18.7} \\
        MERIT  & 18.1         & 17.1          & 19.1          & 20.1          \\
        SPA    & 16.7         & 17.9          & 18.9          & 19.7          \\
        PSAP   & 23.8         & 17.1          & 18.4          & 19.7          \\
        ER     & 15.2         & 17.1          & 18.6          & 19.7          \\
        SNPA   & \textbf{9.6} & 17.9          & 19.0          & 19.9          \\
        VCA    & 15.5         & 17.6          & 19.4          & 20.6          \\
        \hline
        \addlinespace
        \hline
        \multicolumn{5}{c}{Dataset 4}                                         \\
        \hline
               & $\nu = 0.2$  & $\nu = 0.4$   & $\nu =0.6$    & $\nu = 0.8$   \\
        \hline
        EEHT-A & 27.3         & 20.9          & 13.8          & 25.5          \\
        EEHT-B & \textbf{3.5} & 19.2          & 13.8          & 16.2          \\
        EEHT-C & 5.3          & \textbf{7.1}  & \textbf{12.8} & \textbf{14.1} \\
        MERIT  & 8.5          & 13.4          & 27.9          & 17.2          \\
        SPA    & 7.0          & 10.7          & 14.2          & 17.0          \\
        PSPA   & 4.4          & 10.2          & 14.7          & 17.6          \\
        ER     & \textbf{3.5} & 10.7          & 14.7          & 17.5          \\
        SNPA   & 6.1          & 10.7          & 14.2          & 17.0          \\
        VCA    & 22.8         & 26.1          & 13.2          & 17.1          \\
        \hline
    \end{tabular}
\end{table}

\subsection{Hyperspectral Unmixing of the Urban HSI dataset} \label{Subsec: unmixing of Urban}
\begin{figure}[h]
    \centering
    \includegraphics[width=0.6\linewidth]{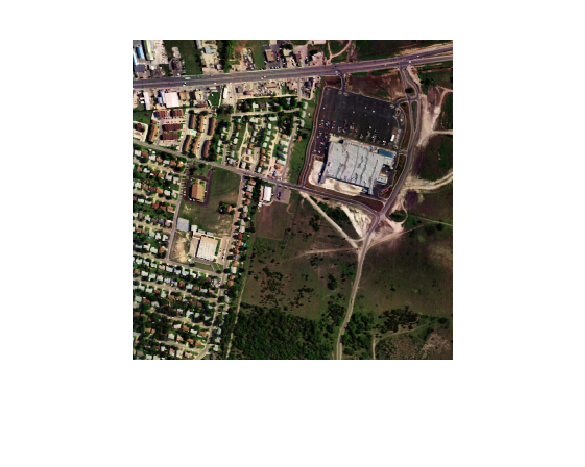}
    \caption{RGB image of Urban.}
    \label{Fig: RGB image of urban}
\end{figure}

Finally, we conducted an experimental study on hyperspectral unmixing of the Urban HSI dataset.
The image was taken over Copperas Cove, Texas, USA by the HYDICE sensor.
It consists of $307 \times 307$ pixels with $210$ bands from 400 nm to 2500 nm.
Following the procedure in \cite{Zhu14, Zhu17}, we removed dirty bands from the image.
The resulting image had 162 clean bands.
The experimental studies in \cite{Zhu14, Zhu17} showed that there are mainly $4$-$6$ endmembers in the image.
Following the settings used in \cite{Gil18}, our experiments set the number of endmembers to $6$:
Asphalt, Grass, Tree, Roof 1, Roof 2 and Soil.
These six endmember signatures were identified in \cite{Zhu17}.
We used them as reference signatures.
Figure \ref{Fig: RGB image of urban} displays the RGB image of the Urban dataset.

We did not achieve satisfactory results for Urban by applying the methods that
were tested on the synthetic HSI datasets described in Section \ref{Subsec: endmember extraction performance}.
The estimated endmember signatures were not sufficiently close to the reference ones.
Consequently, we conducted data-specific preprocessing aimed at enhancing the endmember extraction performance of these methods.
The preprocessing was based on a reasonable assumption that
there are a number of pixels close to each pure pixel in an HSI.
Indeed, this assumption holds in the case of the Urban dataset: see Table \ref{Tab: number of pixels close to each endmember}
that summarizes the number of pixels within an MRSA value of $0.05$ relative to the reference signatures of the endmembers.
Here, we say that pixels are \textit{isolated} if there are almost no pixels close to them.
Under this assumption, even if we remove isolated pixels,
there would be little effect on the number of pure pixels.
It is possible that some of the isolated pixels contain large amounts of noise,
which would worsen the endmember extraction performance of the methods.

\begin{table}[h]
    \centering
    \caption{Number of pixels within an MRSA value of $0.05$ relative to the reference signatures of the endmembers for Urban.}
    \label{Tab: number of pixels close to each endmember}
    \begin{tabular}{c c c c c c}
        \hline
        Asphalt & Grass  & Tree   & Roof 1 & Roof 2 & Soil  \\
        \hline
        939     & 16,435 & 14,122 & 1,493  & 70     & 1,491 \\
        \hline
    \end{tabular}
\end{table}

\begin{figure}[h]
    \centering
    \includegraphics[width=0.75\linewidth]{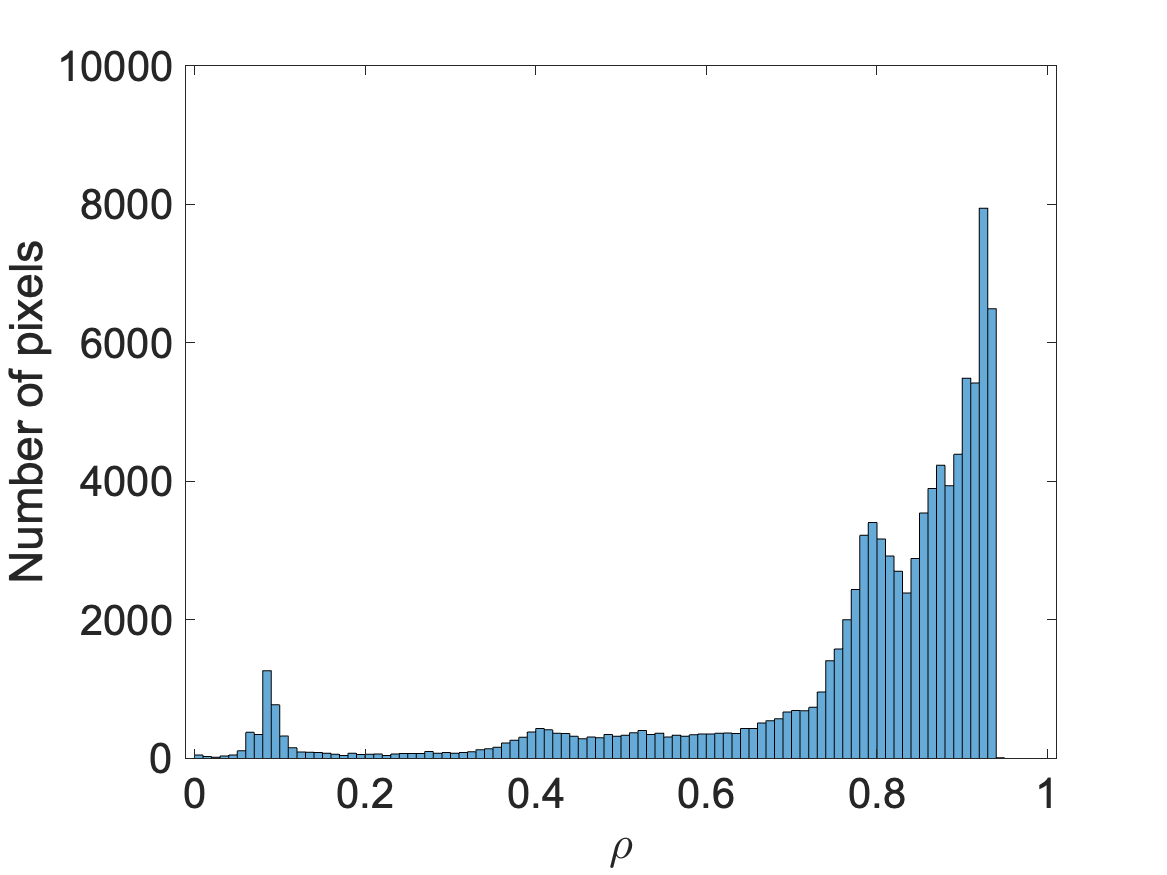}
    \caption{Histogram of neighborhood density $\rho$ versus the number of pixels for Urban with $\phi = 0.4$ (bin size 0.01).}
    \label{Fig: density}
\end{figure}

To find isolated pixels, we introduced the {\it neighborhood density} of each pixel, defined as
\begin{align*}
    \rho(i; \phi) = \frac{1}{n}|\NH(i; \phi)| \in [0,1]
\end{align*}
for the neighborhood of the $i$th pixel $\a_i$ under an MRSA value of $\phi$,
\begin{align*}
    \NH(i; \phi) = \left\{ j \in \NC  \mid \ \MRSA(\a_i, \a_j) \le \phi \right\},
\end{align*}
and the total number $n$ of pixels in the HSI.
For Urban, we examined the neighborhood density $\rho$ while varying $\phi$.
Figure \ref{Fig: density} displays a histogram of $\rho$ with a bin size of 0.01 when $\phi = 0.4$.
We can see from the figure that there is a small peak in the low-density range from $\rho = 0.08$ to $0.09$.
In the experiments, we set $\phi = 0.4$ and removed all pixels $\a_i$ satisfying $\rho(i; \phi) \le \omega$
for a truncation parameter $\omega$ specified in advance.
We ran the same methods as in Section \ref{Subsec: endmember extraction performance} for Urban
while varying $\omega$ from $0$ to $0.15$.

First, let us examine the experiments on Urban with $\phi = 0.4$ and $\omega = 0.1$.
Under these settings, there were $2955$ pixels $\a_i$ ($3\%$ of the total number of pixels) satisfying $\rho(i; \phi) \le \omega$.
We removed all of those pixels.
For EEHT, we set the parameters $\zeta$ and $\eta$ to $(\zeta, \eta) = (50, 300)$.
For MERIT, we set the parameter $\lambda$ to $10^{-6}, 10^{-5}, ..., 10^{3}$,
while fixing the parameter $\mu$ to its default value of $10^{-5}$.
For each $\lambda$ value, we ran MERIT and calculated the MRSA score.
The score was minimized when $\lambda = 10^2$.
Therefore, we report the results of MERIT with $\lambda = 10^2$.

Table \ref{Tab: exp 3 - table of MRSA} summarizes the MRSA values for each endmember and their averages (i.e., the MRSA scores).
We can see from the table that the MRSA values of EEHT-C are smaller than those of the other methods,
except in the case of Roof 2.
Regarding computational time, EEHT-A, -B, and -C each took approximately 10 hours.

\begin{table}[h]
    \centering
    \caption{MRSA values ($\times 10^2$) of the methods for Urban with $\phi = 0.4$ and $\omega = 0.1$.
        The best scores are highlighted in bold.}
    \label{Tab: exp 3 - table of MRSA}
    \begin{tabular}{l r r r r r r  r}
        \hline
               & Asphalt      & Grass        & Tree         & Roof 1       & Roof 2        & Soil         & Avg.         \\
        \hline
        EEHT-A & 11.2         & 15.7         & 44.2         & 12.5         & 17.1          & 10.4         & 18.5         \\
        EEHT-B & 27.4         & 15.7         & 5.8          & 24.8         & 57.5          & 10.4         & 23.6         \\
        EEHT-C & \textbf{9.2} & \textbf{5.7} & \textbf{3.4} & \textbf{7.1} & 18.9          & \textbf{3.2} & \textbf{7.9} \\
        MERIT  & 20.5         & 33.6         & 14.0         & 12.5         & 19.5          & 16.7         & 19.5         \\
        SPA    & 20.5         & 35.4         & 4.7          & 12.5         & \textbf{17.1} & 16.7         & 17.8         \\
        PSPA   & 20.5         & 6.4          & 4.7          & 12.5         & \textbf{17.1} & 16.7         & 13.0         \\
        ER     & 20.5         & 18.4         & 4.7          & 12.5         & 29.2          & 16.7         & 17.0         \\
        VCA    & 16.2         & 41.0         & 15.3         & 13.7         & 50.8          & 31.1         & 28.0         \\
        SNPA   & 12.7         & 27.0         & 4.7          & 12.5         & 44.6          & 16.7         & 19.7         \\
        \hline
    \end{tabular}
\end{table}

\begin{figure}[h]
    \centering
    \includegraphics[width=0.95\linewidth]{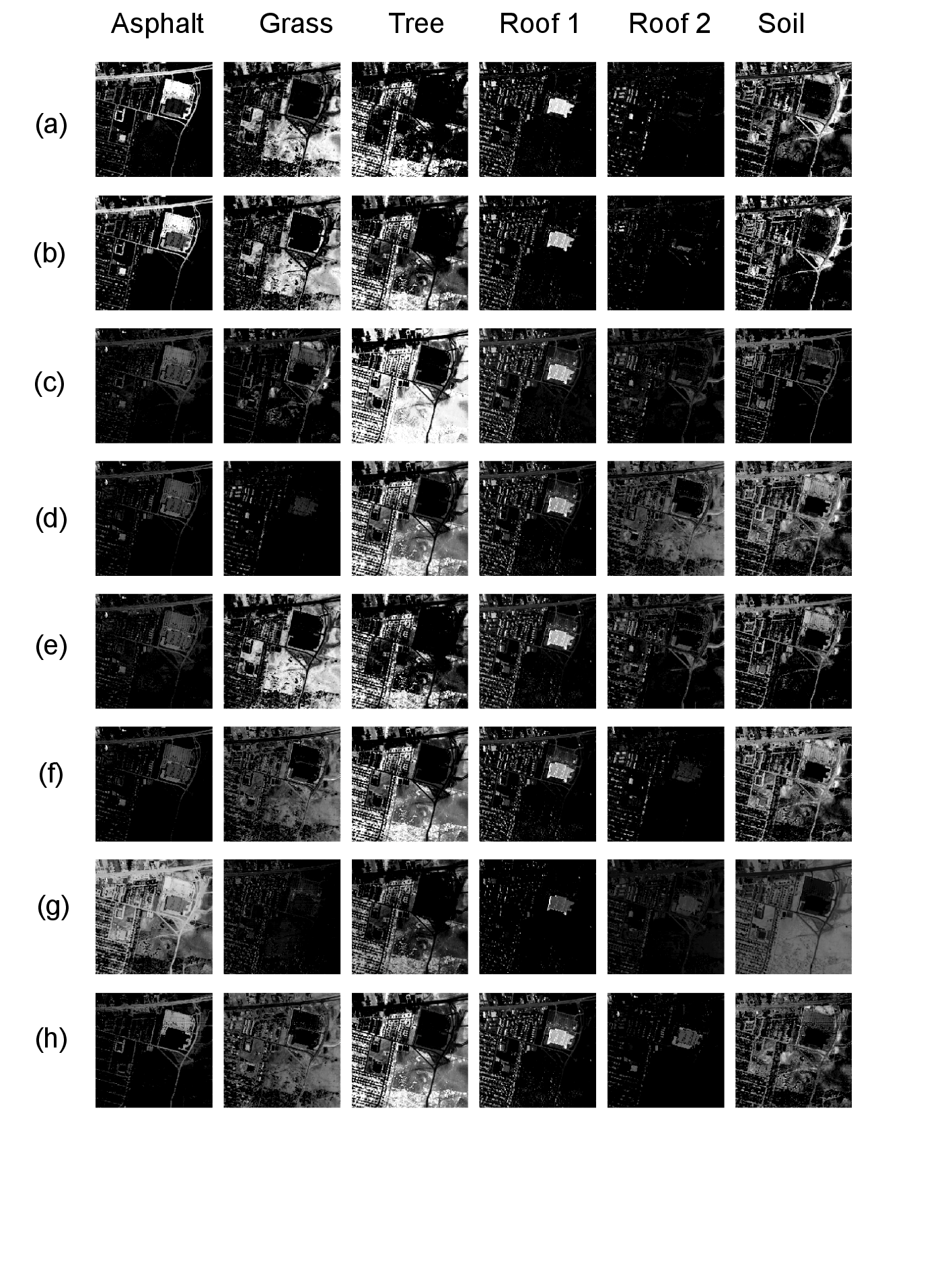}
    \caption{Ground truth and abundance maps obtained by the methods for Urban with $\phi = 0.4$ and $\omega = 0.1$:
        (a) ground truth obtained by using the reference signatures,
        (b) EEHT-C, (c) MERIT, (d) SPA, (e) PSPA, (f) ER, (g) VCA, and (h) SNPA.}
    \label{Fig: exp 3 - abundance maps}
\end{figure}

To assess the effectiveness of EEHT-C further,
we computed abundance maps by solving problem (\ref{Prob: abundance computation})
using the estimated endmember signatures, and compared them with the ground truth derived from the reference signatures.
Figure \ref{Fig: exp 3 - abundance maps} displays the resulting maps.
We can see that the abundance maps obtained by EEHT-C are similar to those obtained by the reference signatures, except Roof 2.
In addition, we quantitatively evaluated the accuracy of the abundance maps.
Let $\hat{\h}_1, \ldots, \hat{\h}_n \in \Real^r$ be the abundance vectors estimated by the methods,
and let $\h_1, \ldots, \h_n \in \Real^r$ be the corresponding ground truth vectors derived from the reference signatures.
The root-mean-square error (RMSE) of the abundance fractions is defined as:
\begin{align*}
    \sqrt{\frac{1}{rn} \sum_{j=1}^{n} \| \hat{\h}_j - \h_j \|_2^2}.
\end{align*}
We also evaluated the reconstruction error (RE) of the unmixing process.
To compute RE,
we first construct the estimated endmember signature matrix $\hat{W} = [\a_{i_1}, \ldots, \a_{i_r}]$
using the columns $\a_{i_1}, \ldots, \a_{i_r}$ of the HSI matrix $A \in \Real^{d \times n}$ returned by the methods.
Each reconstructed column $\hat{\a}_j$ is then defined as $\hat{\a}_j = \hat{W} \hat{\h}_j$.
Given the original columns $\a_1, \ldots, \a_n$ of $A$ and their reconstructions $\hat{\a}_1, \ldots, \hat{\a}_n$,
RE is defined as:
\begin{align*}
    \sqrt{\frac{1}{dn} \sum_{j=1}^{n} \| \hat{\a}_j - \a_j \|_2^2}.
\end{align*}
Table \ref{Tab: exp 3 - quantitative evaluation of abundance maps} summarizes the RMSE and RE values for each method,
indicating that EEHT-C achieved the best RMSE and shared the best RE performance with VCA.

\begin{table}[h]
    \centering
    \caption{RMSE ($\times 10^2$) and RE ($\times 10^2$) values of the abundance maps estimated by each method
        for Urban with $\phi = 0.4$ and $\omega = 0.1$. The best scores are highlighted in bold.}
    \label{Tab: exp 3 - quantitative evaluation of abundance maps}
    \begin{tabular}{l  r r r r r r  r}
        \hline
             & EEHT-C        & MERIT & SPA  & PSPA & ER   & VCA           & SNPA \\
        \hline
        RMSE & \textbf{11.1} & 31.3  & 26.5 & 16.2 & 19.2 & 28.3          & 19.3 \\
        RE   & \textbf{14.6} & 25.3  & 32.7 & 28.5 & 33.9 & \textbf{14.6} & 32.8 \\
        \hline
    \end{tabular}
\end{table}

\begin{figure}[h]
    \centering
    \includegraphics[width=0.75\linewidth]{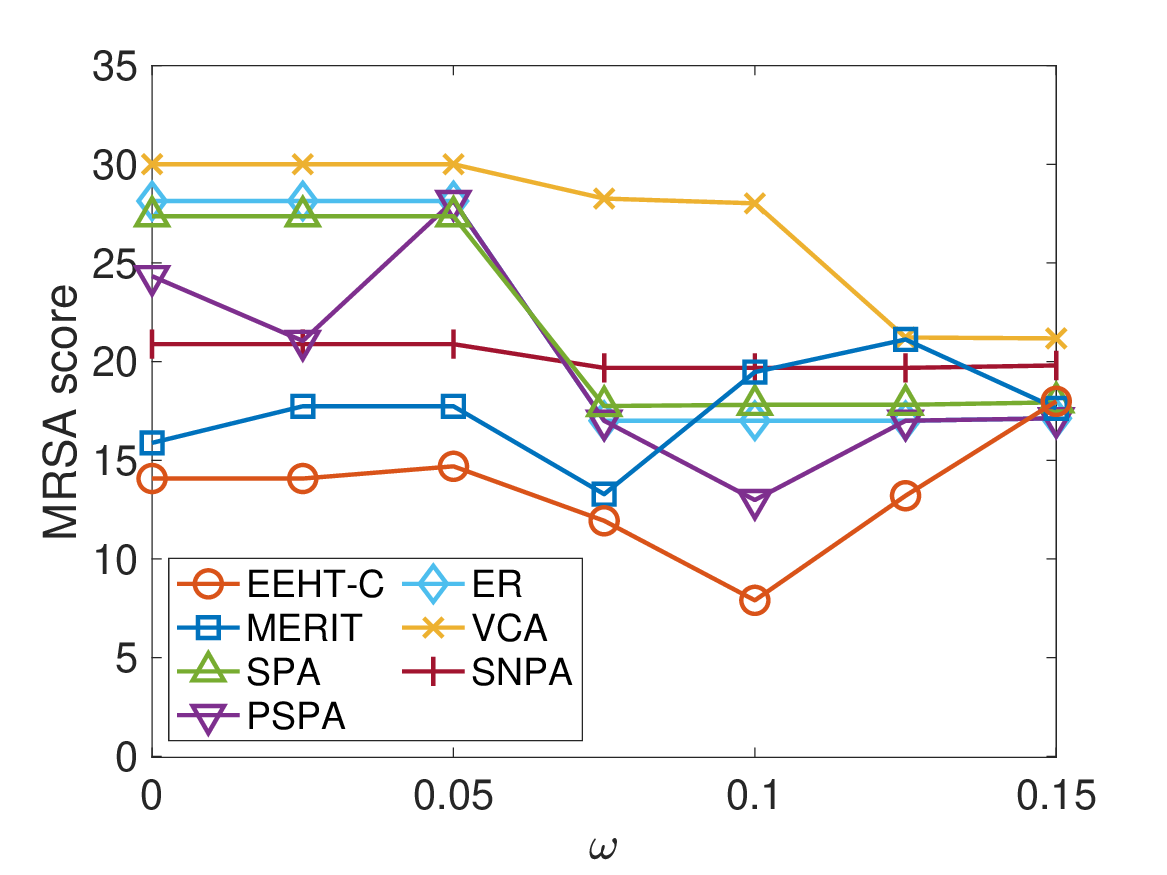}
    \caption{MRSA scores ($\times 10^2$) of methods for Urban with
        $\phi = 0.4$ and varying $\omega$ from $0$ to $0.15$ in increments of $0.025$.}
    \label{Fig: exp 3 - graph of average MRSA}
\end{figure}

Next, let us examine the experiments on Urban with $\phi = 0.4$ and varying $\omega$ from $0$ to $0.15$ in increments of $0.025$.
Figure \ref{Fig: exp 3 - graph of average MRSA} shows the MRSA scores.
We can see from the figure that
EEHT-C achieved better MRSA scores than the other methods, except when $\omega = 0.15$.

The MRSA scores of EEHT-C reached their minimum at $\omega = 0.1$.
When $\omega$ is set to a large value,
some of the pixels $\a_i$ satisfying $\rho(i; \phi) \le \omega $ are no longer isolated and may be close to pure pixels.
Removing such pixels can affect the number of pure pixels in the HSI.
To explore this,
we set $\phi = 0.4$ and $\omega = 0.125$ and removed all pixels $\a_i$ satisfying $\rho(i; \phi) \le \omega $.
At this setting,
no pixels existed with an MRSA value of $0.05$ relative to the reference signature of Roof 1.
Accordingly, when $\omega \ge 0.125$,
it was impossible for EEHT-C to identify pixels containing the spectral signature of Roof 1 with high purity.
For this reason, the  MRSA scores of EEHT-C increased.

As shown in Figure \ref{Fig: exp 3 - graph of average MRSA},
the endmember extraction performance of the methods depends on the choice of $\omega$.
Details of the observation for choosing the value of $\omega$ are provided 
in Appendix \ref{Appx: Effect of parameter omega on endmember extraction performance}.

\section{Concluding Remarks} \label{Sec: concluding remarks}
This study was motivated by the following question:
although the theoretical results shown in \cite{Gil13, Miz22} suggest that
Hottopixx methods are effective for endmember extraction problems, is this really true?
Hottopixx methods require solving computationally expensive LP problems, called Hottopixx models,
which makes them challenging to use in practice.
To address this issue, we developed an efficient and effective implementation of Hottopixx, EEHT,
and provided numerical evidence indicating that
EEHT can estimate the endmember signatures of real HSIs with reasonable accuracy.

We conclude this paper with remarks on potential directions for future research.
Further theoretical investigation is needed to clarify the computational efficiency of RCE,
which is incorporated in EEHT.
In particular, it is important to analyze its iteration complexity
as well as the growth of the index set $\LC$ during the iterations.

As we saw in Section \ref{Subsec: unmixing of Urban},
EEHT can be applied to endmember extraction of HSIs with on the order of one hundred thousand pixels.
However, its computational time is considerably longer than that of greedy methods, such as SPA and its refinements.
To reduce this, we need to focus on the computational cost of RCE,
whose bottleneck lies in solving problems $\Primal(\LC, \LC)$ and $\Dual(\LC, \LC)$ in step 2-1,
since the problem size can be large.
Augmented Lagrangian methods are often employed for solving large-scale LPs; see \cite{Bur06}, for instance.
These methods may help reduce the computational time required by EEHT.

Another promising direction is the use of preprocessing techniques
to remove redundant pixels from HSI data that are far from pure pixels.
Such techniques can substantially reduce the computational burden
of solving $\Primal(\LC, \LC)$ and $\Dual(\LC, \LC)$.
For this approach to be successful,
it is important to develop preprocessing methods that ensure
the remaining pixels still contain pure pixels or those close to them.
We are currently pursuing research along this line and expect to report our findings in the near future.

\bibliographystyle{abbrv}
\bibliography{main}

\appendices

\section{Proof of Theorem \ref{Thm: optimal values of S, P, D}} \label{Appx: Proof of Thm: optimal values of S, P, D}
\begin{IEEEproof}
    \fbox{Part (i)} \
    We show that feasible solutions exist for both $\Primal(\LC, \MC)$ and $\Dual(\LC, \MC)$.
    Setting all variables of $\Dual(\LC, \MC)$ to zero yields a solution that satisfies all of its constraints.
    Thus, $\Dual(\LC, \MC)$ has a feasible solution.
    Moreover, $\FC(\ell, m)$ is nonempty.
    Indeed, let $X \in \Real^{\ell \times m}$ be a matrix defined by
    $X(i,i) = 1$ for $i = 1, \ldots, r$, and $X(i,j) = 0$ for all other indices $(i,j)$.
    Then $X \in \FC(\ell, m)$. By choosing some $\hat{X} \in \FC(\ell, m)$, let $\hat{R} = A(\MC) \Pi - A(\LC) \hat{X}$.
    We show that $(X, F, G, u) = (\hat{X}, \hat{R}^+, \hat{R}^-, \| \hat{R} \|_1)$ is a feasible solution to $\Primal(\LC, \MC)$.
    It is clear that this tuple satisfies the third, fourth, and fifth constraints.
    Regarding the first constraint, we have
    \begin{align*}
        A(\MC) \Pi - A(\LC) X & = A(\MC) \Pi - A(\LC) \hat{X} \\
                              & = \hat{R}                     \\
                              & = \hat{R}^+ - \hat{R}^-       \\
                              & = F - G.
    \end{align*}
    For the second constraint, equality (\ref{Eq: L1 norm expression}) implies that
    \begin{align*}
        \sum_{i=1}^{d} F(i,j) + G(i,j) & = \sum_{i=1}^{d} \hat{R}^+(i,j) + \hat{R}^-(i,j) \\
                                       & \le \|\hat{R}\|_1 = u
    \end{align*}
    holds for each $j = 1, \ldots, m$.
    Thus, $\Primal(\LC, \MC)$ has a feasible solution.
    Accordingly, it follows from the duality theorem that $\opt(\Primal(\LC, \MC)) = \opt(\Dual(\LC, \MC))$.

    \fbox{Parts (ii)-(iv)} \
    Let $(X^*, F^*, G^*, u^*)$ be the optimal solution to $\Primal(\LC, \MC)$.
    Then $X^*$ is a feasible solution to $\HT(\LC, \MC)$.
    Thus, we get $\opt(\HT(\LC, \MC)) \le \| A(\MC)\Pi - A(\LC)X^* \|_1$.
    Furthermore, since $(X^*, F^*, G^*, u^*)$ satisfies all constraints  of $\Primal(\LC, \MC)$,
    we obtain
    \begin{align*}
        \| A(\MC)\Pi - A(\LC)X^* \|_1
         & = \|F^* - G^* \|_1                                              \\
         & = \max_{j=1, \ldots, m} \| F^*(:, j) - G^*(:,j) \|_1            \\
         & \le  \max_{j=1, \ldots, m} \| F^*(:, j) \|_1 + \| G^*(:,j) \|_1 \\
         & = \max_{j=1, \ldots, m} \sum_{i=1}^{d} F^*(i,j) + G^*(i,j)      \\
         & \le u^*  = \opt(\Primal(\LC, \MC)).
    \end{align*}
    Accordingly,
    \begin{align} \label{Ineq: first inequality in the proof of Thm 1}
        \opt(\HT(\LC, \MC)) \le \| A(\MC)\Pi - A(\LC)X^* \|_1 \le \opt(\Primal(\LC, \MC)).
    \end{align}

    Let $\tilde{X}^*$ be the optimal solution to $\HT(\LC, \MC)$,
    and let $\tilde{R} = A(\MC)\Pi - A(\LC)\tilde{X}^*$.
    Then $\tilde{R}$ satisfies the relation $\| \tilde{R} \|_1 = \opt(\HT(\LC, \MC))$.
    As shown in part (i), $(X, F, G, u) = (\tilde{X}^{*}, \tilde{R}^{+}, \tilde{R}^{-}, \| \tilde{R} \|_1)$ is a feasible solution to $\Primal(\LC,\MC)$.
    Accordingly,
    \begin{align} \label{Ineq: second inequality in the proof of Thm 1}
        \opt(\HT(\LC, \MC)) = \| \tilde{R} \|_1  \ge \opt(\Primal(\LC, \MC)).
    \end{align}
    From (\ref{Ineq: first inequality in the proof of Thm 1}) and (\ref{Ineq: second inequality in the proof of Thm 1}),
    we obtain the equalities,
    \begin{align*}
        \opt(\Primal(\LC, \MC)) & =  \opt(\HT(\LC, \MC))                                \\
                                & = \| A(\MC)\Pi - A(\LC)X^* \|_1  = \| \tilde{R} \|_1.
    \end{align*}
    As shown above,
    $X^*$ and $(\tilde{X}^{*}, \tilde{R}^{+}, \tilde{R}^{-}, \| \tilde{R} \|_1)$ are feasible solutions
    to $\HT(\LC, \MC)$ and $\Primal(\LC,\MC)$, respectively.
    Hence, this completes the proof of parts (ii)-(iv).
\end{IEEEproof}

\section{Proof of Theorem \ref{Thm: main}} \label{Appx: Proof of main theorem}
First, we will prove part (i) of Theorem \ref{Thm: main} and then address part (ii).
\begin{IEEEproof}[Proof of part (i) of Theorem \ref{Thm: main}]
    First, we show that $\opt(\HT(\LC, \LC)) \ge  \opt(\HT(\LC, \NC))$ holds under condition (C1).
    The matrix $[X^*, \Gamma^*]$ belongs to $\FC(\ell, n)$, and thus, it is a feasible solution to $\HT(\LC, \NC)$.
    In addition,
    we obtain $\| A(\LC) - A(\LC)X^* \|_1 = \opt(\HT(\LC,\LC))$ by Theorem \ref{Thm: optimal values of S, P, D} (iii),
    and $\|A(\NC \setminus \LC) - A(\LC)\Gamma^*] \|_1 \le \opt(\HT(\LC, \LC))$ by condition (C1) and Theorem \ref{Thm: optimal values of S, P, D} (ii).
    Thus, the objective function value of $\HT(\LC, \NC)$ at $[X^*, \Gamma^*]$ satisfies
    \begin{align*}
        \opt(\HT(\LC, \NC))
         & \le \|A \Pi - A(\LC) [X^*, \Gamma^*] \|_1                            \\
         & = \| \ALN - A(\LC) [X^*, \Gamma^*] \|_1                              \\
         & = \| [A(\LC) - A(\LC)X^*, A(\NC \setminus \LC) - A(\LC)\Gamma^*]\|_1 \\
         & = \opt(\HT(\LC, \LC)).
    \end{align*}

    Next, we show the reverse inequality.
    Let $\hat{X}^* = [\hat{\x}_1^*, \ldots, \hat{\x}_n^*]$ denote the optimal solution to $\HT(\LC, \NC)$.
    We find that
    \begin{align*}
        \opt(\HT(\LC, \NC)) & = \| A \Pi - A(\LC) \hat{X}^* \|_1                                    \\
                            & =  \| \ALN - A(\LC)\hat{X}^* \|_1                                     \\
                            & \ge \| [ A(\LC) - A(\LC) [\hat{\x}_1^*, \ldots, \hat{\x}_\ell^*] \|_1 \\
                            & \ge \opt(\HT(\LC,\LC)).
    \end{align*}
    The last inequality comes from that the matrix $[\hat{\x}_1^*, \ldots, \hat{\x}_\ell^*]$
    belongs to $\FC(\ell, \ell)$ and is, therefore, a feasible solution to $\HT(\LC, \LC)$.
    The two inequalities we showed above yield
    \begin{align*}
        \opt(\HT(\LC, \LC)) = \opt(\HT(\LC, \NC)) = \| A \Pi - A(\LC) [X^*, \Gamma^*] \|_1.
    \end{align*}
    We conclude that $\opt(\HT(\LC, \LC)) = \opt(\HT(\LC, \NC))$ and the matrix $[X^*, \Gamma^*]$
    is the optimal solution to $\HT(\LC, \NC)$.
\end{IEEEproof}

Next, we prove part (ii) of Theorem \ref{Thm: main}.
To do so, we need Lemmas \ref{Lem: claim (a)} and \ref{Lem: claim (b)}.
\begin{itemize}
    \item \textit{Lemma \ref{Lem: claim (a)}:}
          Let $\alpha^*$ be the optimal solution to $\Primal(\LC, \LC)$.
          If $\alpha^*$ satisfies condition (C1), then $\alpha$ can be constructed from $\alpha^*$
          such that $\alpha$ is a feasible solution to $\Primal(\NC, \NC)$,
          and the objective function value of  $\Primal(\NC, \NC)$ at $\alpha$ is $\opt(\Primal(\LC, \LC))$.
    \item \textit{Lemma \ref{Lem: claim (b)}:}
          Let $\beta^*$ be the optimal solution to $\Dual(\LC, \LC)$.
          If $\beta^*$ satisfies condition (C2),
          then $\beta$ can be constructed from $\beta^*$ such that $\beta$ is a feasible solution to $\Dual(\NC, \NC)$,
          and the objective function value of  $\Dual(\NC, \NC)$ at $\beta$ is $\opt(\Dual(\LC, \LC))$.
\end{itemize}
Theorem \ref{Thm: optimal values of S, P, D} (i) tells us that $\opt(\Primal(\LC, \LC)) = \opt(\Dual(\LC, \LC))$.
Accordingly, as we reviewed in Section \ref{Subsec: duality and solution methods for LPs},
the weak duality theorem implies that $\alpha$ and $\beta$ are optimal solutions to $\Primal(\NC,\NC)$ and $\Dual(\NC,\NC)$, respectively.
Furthermore, it means that $\opt(\Primal(\NC, \NC)) = \opt(\Primal(\LC, \LC))$.
As a result, we can establish part (ii) of Theorem \ref{Thm: main}.

Here, we give formal descriptions of Lemmas \ref{Lem: claim (a)} and \ref{Lem: claim (b)} and then prove them.
\begin{lem} \label{Lem: claim (a)}
    Let $\LC \subset \NC$.
    Let $\alpha^* = (X^*, F^*, G^*, u^*)$ be the optimal solution to $\Primal(\LC,\LC)$.
    Assume that condition (C1) holds.
    By letting $\Gamma^* = [\gammab_j^* \mid j \in \NC \setminus \LC]$ for the optimal solution $\gammab_j^*$ to $\AugProb_j(\LC, X^*)$
    and letting $\Pi$ be a permutation matrix of size $n$ satisfying $A\Pi = [A(\LC), A(\NC \setminus \LC)]$,
    construct $\alpha = (X, F, G, u) \in \Real^{n \times n} \times \Real^{d \times n} \times \Real^{d \times n} \times \Real$ as follows:
    \begin{align*}
        X = \Pi
        \begin{bmatrix}
            X^* & \Gamma^* \\
            O   & O
        \end{bmatrix}
        \Pi^\trans, \
        F = R^+,  \ G = R^-  \ \text{and} \ u = \| R \|_1
    \end{align*}
    where $R = A - AX$ for $X$ as constructed above.
    Then, the following hold.
    \begin{enumerate}[(i)]
        \item $\alpha$ is a feasible solution to $\Primal(\NC,\NC)$.
        \item The objective function value of $\Primal(\NC, \NC)$ at $\alpha$ is $\opt(\Primal(\LC, \LC))$,
              i.e., $u = \opt(\Primal(\LC, \LC))$.
    \end{enumerate}
\end{lem}
\begin{IEEEproof}
    As a simplification,
    we can write $X = \Pi \hat{X} \Pi^\trans$  for $X$ above by letting
    \begin{align*}
        \hat{X} =
        \begin{bmatrix}
            X^* & \Gamma^* \\
            O   & O
        \end{bmatrix}.
    \end{align*}

    \fbox{Part (i)} \
    It is sufficient to prove that $\alpha$ satisfies the first, second, and fifth constraints of $\Primal(\NC,\NC)$,
    since $F = R^+ \ge O$ and $G = R^- \ge O$.
    Let us look at the first constraint.
    Since $R$, $F$ and $G$ are constructed as $R = A - AX$, $F = R^+$ and $G = R^-$,
    we have
    \begin{align*}
        A - AX = R = R^+ - R^- = F - G.
    \end{align*}
    Next, let us look at the second constraint. Equality (\ref{Eq: L1 norm expression}) tells us that
    \begin{align*}
        u = \| R \|_1 = \max_{j =1, \ldots, n} \sum_{i=1}^{d} R^+(i,j) + R^-(i,j).
    \end{align*}
    This implies that
    $\sum_{i=1}^{d} F(i,j) + G(i,j) \le u$ holds for $j =1, \ldots, n$.
    Finally, let us look at the fifth constraint.
    Since $X^* \in \FC(\ell, \ell)$ and $\zero \le \gammab_j^* \le \diag(X^*)$,
    we see that $\hat{X} \in \FC(n,n)$.
    Additionally, since $X$ is constructed as $X = \Pi \hat{X} \Pi^\trans$,
    the $(i,j)$th entry of $X$ is given by $\hat{X}(\sigma(i), \sigma(j))$
    for some permutation $\sigma$ of size $n$.
    Hence,  $X \in \FC(n,n)$.
    Thus, the desired result follows.

    \fbox{Part (ii)} \
    The objective function value of $\Primal(\NC, \NC)$ at $\alpha$ can be rewritten as
    \begin{align*}
        u = \| R \|_1 = \| A - AX \|_1
        = \| A\Pi - A\Pi \hat{X} \|_1.
    \end{align*}
    In light of the relation $A\Pi = \ALN$, it can be further expressed as
    \begin{align*}
        \| A\Pi - A \Pi \hat{X} \|_1 = \| [S, T] \|_1
    \end{align*}
    for $S = A(\LC) - A(\LC)X^*$ and $T = A(\NC \setminus \LC) - A(\LC) \Gamma^*$.
    It follows from parts (ii) and (iii) of Theorem \ref{Thm: optimal values of S, P, D} that
    $\opt(\Primal(\LC, \LC)) = \opt(\HT(\LC,\LC))  = \|A(\LC) - A(\LC)X^* \|_1 = \| S \|_1$.
    It follows from condition (C1) that
    $\| T \|_1 = \| A(\NC \setminus \LC) - A(\LC) \Gamma^* \|_1 \le \opt(\Primal(\LC, \LC))$.
    Consequently, we obtain $u = \opt(\Primal(\LC,\LC))$.
\end{IEEEproof}

\begin{lem} \label{Lem: claim (b)}
    Let $\LC \subset \NC$.
    Let $\beta^* = (Y^*, Z^*, \s^*, \t^*, v^*)$	be the optimal solution to $\Dual(\LC,\LC)$.
    Assume that condition (C2) holds.
    By letting $\Pi$ be a permutation matrix of size $n$ satisfying $A\Pi = [A(\LC), A(\NC \setminus \LC)]$,
    and $\Delta = (Y^*)^\trans A(\NC \setminus \LC)$,
    construct $\beta = (Y, Z, \s, \t, v) \in \Real^{d \times n} \times \Real^{n \times n} \times \Real^n \times \Real^n \times \Real$ as follows:
    \begin{align*}
        Y =
        \begin{bmatrix}
            Y^*, O
        \end{bmatrix}
        \Pi^\trans, \
        Z = \Pi
        \begin{bmatrix}
            Z^* & \Delta^+ \\
            O   & O
        \end{bmatrix}
        \Pi^\trans, \
        \s = \Pi
        \begin{bmatrix}
            \s^* \\
            \zero
        \end{bmatrix}, \\
        \t = \Pi
        \begin{bmatrix}
            \t^* \\
            \zero
        \end{bmatrix} \	\text{and} \
        v = v^*.
    \end{align*}
    Then, the following hold.
    \begin{enumerate}[(i)]
        \item $\beta$ is a feasible solution to $\Dual(\NC,\NC)$.
        \item The objective function value of $\Dual(\NC, \NC)$ at $\beta$ is $\opt(\Dual(\LC, \LC))$.
    \end{enumerate}
\end{lem}
\begin{IEEEproof}
    The proof uses the following relation. Let $\a \in \Real^n$ and $P \in \Real^{n \times n}$ be a permutation matrix.
    Then, a straightforward calculation gives
    \begin{align}
        \diag(P \a) = P \diag(\a) P^\trans.
        \label{Eq: diag(P * a)}
    \end{align}
    As a simplification, we can write
    $Y = \hat{Y}\Pi^\trans, Z = \Pi \hat{Z} \Pi^\trans, \s = \Pi \hat{\s}$ and $\t = \Pi \hat{\t}$
    for $\hat{Y}, \hat{Z}, \hat{\s}$ and $\hat{\t}$ above by letting
    \begin{align*}
        \hat{Y} =
        \begin{bmatrix}
            Y^*, O
        \end{bmatrix}, \
        \hat{Z} =
        \begin{bmatrix}
            Z^* & \Delta^+ \\
            O   & O
        \end{bmatrix}, \
        \hat{\s} =
        \begin{bmatrix}
            \s^* \\
            \zero
        \end{bmatrix} \ \text{and} \
        \hat{\t} =
        \begin{bmatrix}
            \t^* \\
            \zero
        \end{bmatrix}.
    \end{align*}

    \fbox{Part (i)} \
    It is sufficient to prove that $\beta$ satisfies the first and the second constraints of $\Dual(\NC, \NC)$,
    since, obviously, it satisfies the other constraints.
    First, look at the second constraint.
    We find that $\beta$ satisfies the second constraint of $\Dual(\NC,\NC)$ if and only if $\beta^*$ satisfies that of $\Dual(\LC, \LC)$.
    Indeed,
    \begin{align*}
                  & -J \cdot \diag(\s) \le Y \le J \cdot \diag(\s)                                                                         \\
        \equivSym & -J \cdot \diag(\Pi \hat{\s}) \le \hat{Y}\Pi^\trans \le J \cdot \diag(\Pi \hat{\s})                                     \\
        \equivSym & -J \Pi \cdot \diag(\hat{\s}) \cdot \Pi^\trans \le \hat{Y}\Pi^\trans \le J \Pi \cdot \diag(\hat{\s}) \cdot \Pi^\trans \
        (\text{by (\ref{Eq: diag(P * a)})})                                                                                                \\
        \equivSym & -J \cdot \diag(\hat{\s}) \le \hat{Y} \le J \cdot \diag(\hat{\s})                                                       \\
        \equivSym & -J \cdot \diag(\s^*) \le Y^* \le J \cdot \diag(\s^*).
    \end{align*}
    Next, look at the first constraint.
    By plugging in $\beta$ on the left-hand side of the constraint,
    we get the following matrix $M \in \Real^{n \times n}$:
    \begin{align*}
        M = A^\trans Y + vI - \diag(\t) - Z^\trans + \diag(Z^\trans \one).
    \end{align*}
    Then, we rewrite $M$ as
    \begin{align*}
        M & = A^\trans \hat{Y} \Pi^\trans + v^* I - \Pi \cdot \diag(\hat{\t}) \cdot \Pi^\trans - \Pi \hat{Z}^\trans \Pi^\trans          \\
          & \qquad + \Pi \cdot \diag(\hat{Z}^\trans \one) \cdot \Pi^\trans \ \text{(by (\ref{Eq: diag(P * a)}))}                        \\
          & = \Pi (A\Pi)^\trans \hat{Y} \Pi^\trans + v^* I - \Pi \cdot \diag(\hat{\t}) \cdot \Pi^\trans - \Pi \hat{Z}^\trans \Pi^\trans \\
          & \qquad + \Pi \cdot \diag(\hat{Z}^\trans \one) \cdot \Pi^\trans,
    \end{align*}
    which yields
    \begin{align*}
        \Pi^\trans M \Pi = (A\Pi)^\trans \hat{Y} + v^* I - \diag(\hat{\t}) - \hat{Z}^\trans + \diag(\hat{Z}^\trans \one).
    \end{align*}
    Here,
    \begin{align*}
        (A\Pi)^\trans \hat{Y}      & =
        \begin{bmatrix}
            A(\LC)^\trans Y^*               & O \\
            A(\NC \setminus \LC)^\trans Y^* & O
        \end{bmatrix}, \\
        \diag(\hat{\t})            & =
        \begin{bmatrix}
            \diag(\t^*) & O \\
            O           & O
        \end{bmatrix},                     \\
        \hat{Z}^\trans             & =
        \begin{bmatrix}
            (Z^*)^\trans      & O \\
            (\Delta^+)^\trans & O
        \end{bmatrix},               \\
        \diag(\hat{Z}^\trans \one) & =
        \begin{bmatrix}
            \diag((Z^*)^\trans \one) & O                             \\
            O                        & \diag((\Delta^+)^\trans \one)
        \end{bmatrix}.
    \end{align*}
    We set $\tilde{M} = \Pi^\trans M \Pi$ and partition $\tilde{M}$ into four blocks
    $\tilde{M}_{11} \in \Real^{\ell \times \ell}, \tilde{M}_{12} \in \Real^{\ell \times (n-\ell)}, \tilde{M}_{21} \in \Real^{(n - \ell) \times \ell} $
    and $\tilde{M}_{22} \in \Real^{(n-\ell) \times (n-\ell)}$
    in the following manner:
    \begin{align*}
        \tilde{M} =
        \begin{bmatrix}
            \tilde{M}_{11} & \tilde{M}_{12} \\
            \tilde{M}_{21} & \tilde{M}_{22}
        \end{bmatrix}
    \end{align*}
    where each block is given as
    \begin{align*}
        \tilde{M}_{11} & = A(\LC)^\trans Y^* + v^* I - \diag(\t^*) - (Z^*)^\trans + \diag( (Z^*)^\trans \one ), \\
        \tilde{M}_{12} & = O,                                                                                   \\
        \tilde{M}_{21} & = A(\NC \setminus \LC)^\trans Y^*  - (\Delta^+)^\trans,                                \\
        \tilde{M}_{22} & = v^* I + \diag( (\Delta^+)^\trans \one).
    \end{align*}
    We observe the following:
    $\tilde{M}_{11} \le O$ since $\beta^*$ satisfies the first constraint of $\Dual(\LC, \LC)$;
    $\tilde{M}_{21} \le O$ since
    \begin{align*}
        \tilde{M}_{21} = A(\NC \setminus \LC)^\trans Y^*  - (\Delta^+)^\trans = \Delta^\trans - (\Delta^+)^\trans \le O;
    \end{align*}
    and, $\tilde{M}_{22} \le O$ since
    \begin{align*}
        \tilde{M}_{22}
         & = v^* I + \diag( (\Delta^+)^\trans \one)                                                              \\
         & = \diag \left( v^* + \one^\trans \Delta^+(:,1), \ldots,  v^* + \one^\trans \Delta^+(:, n-\ell)\right) \\
         & \le O \ (\text{by condition (C2)}).
    \end{align*}
    Accordingly, $\tilde{M} = \Pi^\trans M \Pi \le O \equivSym M \le O$,
    which means that $\beta$ satisfies the first constraint of $\Dual(\NC, \NC)$.
    Consequently, we obtain the desired result.

    \fbox{Part (ii)} \
    By a straightforward calculation, we find that
    \begin{align*}
        \langle A, Y \rangle + r  v - \one^\trans \t
         & = \langle A \Pi, Y \Pi \rangle + r  v - \one^\trans \t    \\
         & = \langle \ALN, [Y^*, O] \rangle + r  v^*                 \\
         & \qquad - \one^\trans \Pi \hat{\t}                         \\
         & = \langle A(\LC), Y^*  \rangle + r v^* - \one^\trans \t^* \\
         & = \opt(\Dual(\LC, \LC)).
    \end{align*}
\end{IEEEproof}

Now, we are ready to prove part (ii) of Theorem \ref{Thm: main}.
\begin{IEEEproof}[Proof of part (ii) of Theorem \ref{Thm: main}]
    Let $\alpha$ and $\beta$ be as constructed in Lemmas \ref{Lem: claim (a)} and \ref{Lem: claim (b)}.
    From the lemmas and Theorem \ref{Thm: optimal values of S, P, D} (i), we find that
    $\alpha$ and $\beta$ are feasible solutions to $\Primal(\NC, \NC)$ and $\Dual(\NC, \NC)$
    and the objective function value of $\Primal(\NC, \NC)$ at $\alpha$ is equal to that of $\Dual(\NC, \NC)$ at $\beta$.
    Thus, the weak duality theorem implies that $\alpha$ and $\beta$ are optimal solutions to $\Primal(\NC, \NC)$ and $\Dual(\NC,\NC)$.
    Furthermore, this means that $\opt(\Primal(\LC, \LC)) = \opt(\Primal(\NC, \NC))$.
    Consequently, it follows from parts (ii) and (iii) of Theorem \ref{Thm: optimal values of S, P, D}
    that $\opt(\HT(\LC, \LC)) = \opt(\HT(\NC, \NC))$ holds and $X$ of $\alpha$ is the optimal solution to $\HT(\NC,\NC)$.
\end{IEEEproof}

\section{Proof of Lemma~\ref{Lem: sign of v}} \label{Appx: Proof of sign of v}
\begin{IEEEproof}
    We plug in the optimal solution to $\Dual(\LC, \LC)$ for the first constraint
    and then express the resulting inequality as $M \le O$ by letting
    \begin{align*}
        M = A(\LC)^\trans Y^* + v^* I - \diag(\t^*) - (Z^*)^\trans + \diag((Z^*)^\trans \one).
    \end{align*}
    The trace of $M$ is given as
    \begin{align*}
        \trace(M) = \langle A(\LC), Y^* \rangle + \ell v^*  - \one^\trans \t^* + \sum_{i \neq j} Z^*(i,j).
    \end{align*}
    Here, the sum of $Z^*(i,j)$ runs over all pairs $(i,j)$ except $i = j$.
    Let $c = \langle A(\LC), Y^* \rangle + \ell v^*  - \one^\trans \t^*$.
    Then, $\opt(\Dual(\LC, \LC)) - c = (r - \ell)v^*$. We use this relation to prove the lemma.
    We have $c \le 0$, since
    \begin{align*}
        c = \langle A(\LC), Y^* \rangle + \ell v^*  - \one^\trans \t^* =  \trace(M) - \sum_{i \neq j} Z^*(i,j) \le 0
    \end{align*}
    by $M \le O$ and $Z^* \ge O$.
    We also have $\opt(\Dual(\LC,\LC)) \ge 0$.
    Indeed, $\opt(\Dual(\LC, \LC)) = \opt(\HT(\LC,\LC))$ by parts (i) and (ii) of Theorem 1, and
    $\opt(\HT(\LC,\LC)) \ge 0$, since the objective function value of $\HT(\LC, \LC)$ takes nonnegative values.
    We thus get
    \begin{align*}
        \opt(\Dual(\LC,\LC)) - c \ge 0 \equivSym (r - \ell) v^* \ge 0.
    \end{align*}
    This implies that $v^* \le 0$,
    since $\LC$ is chosen to be $\ell \ge r \equivSym r - \ell \le 0$.
\end{IEEEproof}

\section{Cluster Construction in Step 2-1 of Algorithm 1} 
\label{Appx: Cluster construction in step 2-1 of Algorithm 1}
We describe the notation used for constructing a cluster $\SC_\ell$ in step 2-1 of Algorithm 1.
We choose a column $\a_i$ of $A$ and sort the columns $\a_1, \ldots, \a_n$ of $A$ by their distance to $\a_i$
in ascending order so that
\begin{align*}
    \| \a_i - \a_{u_1} \|_1 \le \| \a_i - \a_{u_2} \|_1 \le \cdots \le \| \a_i - \a_{u_{n-1}} \|_1
\end{align*}
where $\{i, u_1, \ldots, u_{n-1}\} = \NC$. We then construct
\begin{align*}
    \Omega_i = \{\{i\}, \{i, u_1\}, \{i, u_1, u_2\}, \ldots, \{i, u_1, u_2, \ldots, u_{n-1}\}\}.
\end{align*}
Let $\SC \in \Omega_i$ and $\p \in \Real^n_+$.
We regard $\p$ as a point list for $\SC$ and define the score of $\SC$ with respect to the point list $\p$ as
\begin{align*}
    \score(\SC, \p) = \sum_{u \in \SC} \p(u).
\end{align*}
Additionally, we define the diameter of $\SC$ in $\Omega_i$ by
\begin{align*}
    \diam(\SC) = \max_{u \in \SC} \|\a_i - \a_u \|_1.
\end{align*}
Now, let $\Psi(\p) = \Psi_1(\p) \cup \cdots \cup \Psi_n(\p)$ where
\begin{align*}
    \Psi_i(\p) = \left\{ \SC \in \Omega_i \ \middle|  \ \score(\SC, \p) > \frac{r}{r+1} \right\}.
\end{align*}
Step 2-1 of Algorithm 1 then constructs $\SC_\ell$ as follows:
\begin{align*}
    \SC_\ell = \arg \min_{\SC \in \Psi(\p_\ell)} \diam(\SC).
\end{align*}

\section{Effect of Parameters $\zeta$ and $\eta$ on the Computational Time of RCE-SR} 
\label{Appx: Effect of parameters zeta and eta on elapsed time of RCE-SR}
We conducted experiments to investigate
the effect of the parameters $\zeta$ and $\eta$ on the computational time of RCE-SR.
In these experiments, $\zeta$ was varied from $5$ to $20$ in increments of $5$,
and $\eta$ was varied from $50$ to $200$ in increments of $50$, resulting in $16$ combinations of $(\zeta, \eta)$.
For each combination of $(\zeta, \eta)$, RCE-SR was applied to six datasets,
comprising the five datasets used in the previous experiments and an additional dataset with $n=5000$.
For each dataset, the elapsed time of RCE-SR for $(\zeta, \eta) = (10, 100)$ was used as the baseline
and set to $1$, and the times for the other combinations were evaluated relative to this baseline.
Table \ref{Tab: exp 1 - effect of parameters zeta and eta on elapsed time of RCE-SR} summarizes the results,
showing that even when $(\zeta, \eta)$ were varied across the $16$ combinations,
the elapsed time of RCE-SR remained within the range of $0.8$ to $1.8$.

\begin{table*}[t!]
    \centering
    \caption{Minimum and maximum elapsed times of RCE-SR,
        normalized by the baseline $(\zeta, \eta) = (10, 100)$,
        for the $16$ combinations of $(\zeta, \eta)$.}
    \label{Tab: exp 1 - effect of parameters zeta and eta on elapsed time of RCE-SR}
    \begin{tabular}{l  c c c c c c}
        \hline
            & $n=500$ & $n=1000$ & $n=1500$ & $n=2000$ & $n=2500$ & $n=5000$ \\
        \hline
        Min & 0.96    & 0.84     & 0.82     & 0.92     & 0.98     & 1.01     \\
        Max & 1.72    & 1.40     & 1.38     & 1.48     & 1.62     & 1.22     \\
        \hline
    \end{tabular}
\end{table*}

\begin{table*}[t!]
    \centering
    \caption{MRSA scores ($\times 10^2$) on Urban,
        obtained by varying $\phi$ and choosing $\omega$ according to criteria (1) and (2).
        The best scores are highlighted in bold.}
    \label{Tab: exp 3 - table of average MRSA with varying phi}
    \begin{tabular}{l  r r r r r r r}
        \hline
        $(\phi ,\omega)$ & EEHT-C       & MERIT & SPA  & PSPA & ER   & VCA  & SNPA \\
        \hline
        (0.40, 0.10)     & \textbf{7.9} & 19.5  & 17.8 & 13.0 & 17.0 & 28.0 & 19.7 \\
        (0.45, 0.15)     & \textbf{8.4} & 17.0  & 17.8 & 17.0 & 17.0 & 28.3 & 19.7 \\
        (0.50, 0.30)     & \textbf{9.3} & 17.0  & 17.8 & 17.0 & 17.0 & 28.3 & 19.7 \\
        (0.55, 0.45)     & \textbf{8.2} & 11.9  & 17.8 & 17.0 & 17.0 & 28.0 & 19.7 \\
        (0.60, 0.60)     & \textbf{8.2} & 17.0  & 17.8 & 17.0 & 17.0 & 28.3 & 19.7 \\
        \hline
    \end{tabular}
\end{table*}

\section{Effect of Parameter $\omega$ on Endmember Extraction Performance} 
\label{Appx: Effect of parameter omega on endmember extraction performance}
Our data-specific preprocessing for the Urban HSI dataset requires us to choose two parameters $\phi$ and $\omega$.
In particular, as shown in Figure \ref{Fig: exp 3 - graph of average MRSA},
the endmember extraction performance of the methods depends on the choice of $\omega$.
When plotting the histogram of density $\rho$ versus the number of pixels while varying $\phi$ from $0.45$ to $0.6$ in increments of $0.05$,
as in the case where $\phi = 0.4$, a small peak appeared in a low-density range.
Therefore, on the basis of the experimental results for $\phi = 0.4$,
we can describe the criteria for choosing $\omega$ when $\phi$ is between $0.45$ and $0.6$ as follows.
\begin{enumerate}[(1)]
    \item When drawing the histogram of density $\rho$, a peak in a low-density range should be truncated by $\omega$.
    \item The value of $\omega$ should be small to prevent pixels close to pure pixels from being removed.
\end{enumerate}

We chose $\omega$ by following criteria (1) and (2) and ran EEHT-C and six existing methods on Urban.
Table \ref{Tab: exp 3 - table of average MRSA with varying phi} summarizes the MRSA scores.
We see that the MRSA scores of EEHT-C are nearly unchanged for $\phi$ from $0.4$ to $0.6$.
These results demonstrate that criteria (1) and (2) for choosing $\omega$ yield
consistent endmember extraction results for EEHT-C on Urban when $\phi$ belongs to this range.

\end{document}